\documentclass[aps,prd,preprint]{revtex4}
\usepackage{graphicx, bm}
\usepackage{subfig}
\usepackage{float}
\begin{document}

\title{Model independent study for the anomalous quartic $WW\gamma\gamma$ couplings at Future Electron-Proton Colliders}

\author{V. Ari}
\email[]{vari@science.ankara.edu.tr} \affiliation{Department of Physics, Ankara University, Turkey}

\author{E. Gurkanli}
\email[]{egurkanli@sinop.edu.tr}\affiliation{Department of Physics, Sinop University, Turkey}

\author{A. A. Billur}
\email[]{abillur@cumhuriyet.edu.tr} \affiliation{Deparment of Physics, Sivas Cumhuriyet University, Turkey}

\author{M. Koksal}
\email[]{mkoksal@cumhuriyet.edu.tr} \affiliation{Department of Optical Engineering, Sivas Cumhuriyet University, Turkey}

\begin{abstract}

The Large Hadron Electron Collider and the Future Circular Collider-hadron electron with high center-of-mass energy and luminosity allow to better understand the Standard Model and to examine new physics beyond the Standard Model in the electroweak sector. Multi-boson processes permit for a measurement of the gauge boson self-interactions of the Standard Model that can be used to determine the anomalous gauge boson couplings. For this purpose, we present a study of the process $ep \rightarrow \nu_{e} \gamma \gamma j$ at the Large Hadron Electron Collider with center-of-mass energies of 1.30, 1.98 TeV and at the Future Circular Collider-hadron electron with center-of-mass energies of 7.07, 10 TeV to interpret the anomalous quartic $WW\gamma\gamma$ gauge couplings using a model independent way in the framework of effective field theory. We obtain the sensitivity limits at $95\%$ Confidence Level on 13 different anomalous couplings arising from dimension-8 operators.
The best limit in $f_{Mi}/\Lambda^{4}$ ($i=0,1,2,3,4,5,7$) parameters is obtained for $f_{M2}/\Lambda^{4}$ parameter while the best sensitivity derived on $f_{Ti}/\Lambda^{4}$ ($i=0,1,2,5,6,7$) parameters  is obtained for $f_{T5}/\Lambda^{4}$. In addition, this study is the first report on the anomalous quartic couplings determined by effective Lagrangians at $ep$ colliders.
\end{abstract}

\maketitle

\section{Introduction}

New physics beyond the Standard Model (SM) refers to the theoretical developments needed to clarify the deficiencies of the SM, such as neutrino oscillations, matter-antimatter asymmetry, the origin of mass, the strong CP problem and the nature of dark energy and dark matter. For this reason, new physics models beyond the SM are investigated in various processes at colliders. One of the ways to research new physics models is to study the anomalous gauge boson couplings. The triple and quartic gauge boson couplings that define the strengths of the gauge boson self-interactions are exactly determined by the electroweak $SU(2)\times U(1)$ gauge symmetry of the SM. The triple and quartic gauge boson couplings contribute directly to multi-boson production in the final state of the examined processes at colliders and the precise measurements of the processes involving these couplings can further confirm the SM. Moreover, possible deviations the triple and quartic gauge boson couplings of the SM may be a proof of new physics beyond the SM.

An $ep$ collider may be a good idea to complement the LHC physics program and to investigate for possible effects of new physics beyond the SM.
By precisely analyzing the interactions of the quartic gauge bosons, we may detect the effects of the possible new physics in these colliders. The envisaged future  $ep$ colliders are the Large Hadron electron Collider (LHeC) \cite{lhec} and Future Circular Collider-hadron electron (FCC-he) \cite{fcc}. These colliders are designed to generate $ep$ collisions at center-of-mass energies from 1.30 TeV to 10 TeV. The LHeC has an integrated luminosity of $100$ fb$^{-1}$ and it is planned to collide electron beams with an energy from 60 GeV to possibly 140 GeV with 7 TeV proton beams. However, the FCC-he mode is projected to be realized by accelerating electrons up to 500 GeV and colliding them with the proton beams at the energy of 50 TeV.

One consideration when examining the anomalous quartic boson couplings is to isolate only one of these quartic couplings. For example, an important advantage of the process $pp \rightarrow p\gamma^{*}\gamma^{*}p\rightarrow pWWp$ \cite{de1} through the subprocess $\gamma^{*} \gamma^{*}\rightarrow WW$ at the LHC is that it isolates $WW\gamma\gamma$ coupling from the other quartic couplings. In addition, we can easily see that the process $ep \rightarrow \nu_{e}\gamma\gamma j$ isolates the $WW\gamma\gamma$ coupling.

Photons in the final state of the process $ep \rightarrow \nu_{e}\gamma \gamma j$ at the LHeC and FCC-he have the advantage of being identifiable with high purity and efficiency. The diphoton channels are especially sensitive for new physics beyond the SM in terms of modest backgrounds, excellent mass resolution and the clean experimental signature.

For these purposes, motivated by the comprehensive physical program of the LHeC and FCC-he, we carry out a work to examine the effects to the total cross section of the process $ep \rightarrow \nu_{e}\gamma\gamma j$ of the anomalous quartic $WW\gamma\gamma$ couplings determined by dimension-8 operators. For the investigation, we consider the LHeC's center-of-mass energies of $1.30,1.98$ TeV and integrated luminosity of 100 fb$^{-1}$, as well as the FCC-he's center-of-mass energies of $7.07,10$ TeV and integrated luminosity of 1000 fb$^{-1}$.

Therefore, the context of this study is planned as follows: In Section II, we introduce effective interactions for the anomalous quartic $WW\gamma\gamma$ couplings. In Section III, we perform the numerical analysis of the process $ep \rightarrow \nu_{e} \gamma \gamma j$ at the LHeC and FCC-he to obtain limits on the anomalous quartic couplings. Nevertheless, we discuss conclusions in final Section.

\section{THE EFFECTIVE LAGRANGIAN FOR THE ANOMALOUS QUARTIC $WW\gamma\gamma$ COUPLINGS}

Possible deviations arising from new physics for the triple and quartic gauge boson couplings in the electroweak sector can be parameterized in a model independent framework by means of the effective Lagrangian method. In the literature, the anomalous quartic gauge boson couplings are defined by either linear or non-linear effective Lagrangians. First, the non-linear effective Lagrangians can be used to determine possible deviations from the SM by introducing the anomalous quartic gauge boson couplings via dimension-6 operators. Before the discovery of the Higgs boson at the LHC, these Lagrangians were formed by a nonlinear representation of the spontaneously broken gauge symmetry, considering that there is no Higgs boson in the low energy spectrum. For the anomalous quartic $WW\gamma\gamma$ couplings, the non-linear effective Lagrangians that conserve charge conjugation and parity are determined by \cite{gg}

\begin{eqnarray}
\mathcal{L}_{eff}=\mathcal{L}_{0}+\mathcal{L}_{c}
\end{eqnarray}

\begin{eqnarray}
\mathcal{L}_{0}=-\frac{\pi \alpha}{4}\frac{a_{0}}{\Lambda^{2}}F_{\mu\nu} F^{\mu\nu} W_{\alpha}^{(i)}W^{\alpha (i)}
\end{eqnarray}

\begin{eqnarray}
\mathcal{L}_{c}=-\frac{\pi \alpha}{4}\frac{a_{c}}{\Lambda^{2}}F_{\mu\alpha} F^{\mu\beta} W^{\alpha(i)}W^{(i)}_{\beta}
\end{eqnarray}
where $W^{\alpha(i)} (i=1,2,3)$ is the $SU(2)_{W}$ triplet and $F_{\mu\nu}=\partial_{\mu}A_{\nu}-\partial_{\nu}A_{\mu}$ represents the electromagnetic field tensor, $a_{0}$ and $a_{c}$ are the anomalous coupling parameters.

Dimension-8 operators are described by using a linear representation of the spontaneously broken gauge symmetry of the SM. In this case, the anomalous quartic gauge boson couplings are built by extending the SM Lagrangian with terms including dimension-8 operators as this is the lowest dimension that defines the quartic gauge boson couplings without exhibiting triple gauge-boson couplings \cite{co}. Therefore, the linear effective Lagrangians can be given as follows

\begin{eqnarray}
\mathcal{L}_{eff}=\sum_{j=1,2}\frac{f_{Sj}}{\Lambda^{4}}\uppercase{o}_{Sj}+\sum_{j=0,1,2,5,6,7,8,9}\frac{f_{Tj}}{\Lambda^{4}}\uppercase{o}_{Tj}+\sum_{j=0}^{7}\frac{f_{Mj}}{\Lambda^{4}}\uppercase{o}_{Mj}.
\end{eqnarray}
There are 17 different operators that define the anomalous quartic gauge boson couplings. The indices \textit{S}, \textit{T} and \textit{M} of the couplings represent three class operators.

The first class of these operators, two independent operators including covariant derivative of the Higgs doublet are generated by

\begin{eqnarray}
\uppercase{o}_{S0}&=&[(D_{\mu}\Phi)^{\dagger} D_{\nu}\Phi] \times [(D^{\mu}\Phi)^{\dagger} D^{\nu}\Phi],\\
\uppercase{o}_{S1}&=&[(D_{\mu}\Phi)^{\dagger}D^{\mu}\Phi]\times [(D_{\nu}\Phi)^{\dagger}D^{\nu}\Phi].
\end{eqnarray}
The $\uppercase{o}_{S0}$ and $\uppercase{o}_{S1}$ operators involve quartic $WWWW$, $WWZZ$ and $ZZZZ$ couplings.

Seven operators in second class derive the anomalous quartic gauge boson couplings that are obtained by considering two electroweak field strength tensors and two covariant derivatives of the Higgs doublet

\begin{eqnarray}
\uppercase{o}_{M0}&=&Tr[W_{\mu\nu}W^{\mu\nu}]\times[(D_{\beta}\Phi)^{\dagger}D^{\beta}\Phi],\\
\uppercase{o}_{M1}&=&Tr[W_{\mu\nu}W^{\nu\beta}]\times[(D_{\beta}\Phi)^{\dagger}D^{\mu}\Phi],\\
\uppercase{o}_{M2}&=&Tr[B_{\mu\nu}B^{\mu\nu}]\times[(D_{\beta}\Phi)^{\dagger}D^{\beta}\Phi],\\
\uppercase{o}_{M3}&=&Tr[B_{\mu\nu}B^{\nu\beta}]\times[(D_{\beta}\Phi)^{\dagger}D^{\mu}\Phi],\\
\uppercase{o}_{M4}&=&[(D_{\mu}\Phi)^{\dagger}W_{\beta\nu} D^{\mu}\Phi]\times B^{\beta\nu},\\
\uppercase{o}_{M5}&=&[(D_{\mu}\Phi)^{\dagger}W_{\beta\nu} D^{\nu}\Phi]\times B^{\beta\mu},\\
\uppercase{o}_{M6}&=&[(D_{\mu}\Phi)^{\dagger}W_{\beta\nu}W^{\beta\nu} D^{\mu}\Phi],\\
\uppercase{o}_{M7}&=&[(D_{\mu}\Phi)^{\dagger}W_{\beta\nu}W^{\beta\mu} D^{\nu}\Phi].
\end{eqnarray}
Here, the field strength tensors of $W_{\mu\nu}$ and $B_{\mu\nu}$ gauge fields are expressed as
\begin{eqnarray}
W_{\mu\nu}&=&\frac{i}{2} g \tau^i ( \partial_{\mu} W_{\nu}^i-\partial_{\nu}W_{\mu}^i+g \epsilon_{ijk} W_{\mu}^j W_{\nu}^k),\nonumber\\
B_{\mu\nu}&=&\frac{i}{2}g'(\partial_{\mu}B_{\nu}-\partial_{\nu}B_{\mu}).
\end{eqnarray}
where $\tau^i (i=1,2,3)$ shows the $SU(2)$ generators, $g=e/sin\theta_W$, $g'=g/cos \theta_W$, $e$ and $\theta_W$ are the unit of electric charge and the Weinberg angle, respectively.

The final class have 8 operators that consist of four field strength tensors. These operators generate the following quartic anomalous couplings:

\begin{eqnarray}
\uppercase{o}_{T0}&=& Tr[W_{\mu\nu}W^{\mu\nu}]\times [W_{\alpha\beta}W^{\alpha\beta}],\\
\uppercase{o}_{T1}&=& Tr[W_{\alpha\nu}W^{\mu\beta}]\times [W_{\mu\beta}W^{\alpha\nu}],\\
\uppercase{o}_{T2}&=&Tr[W_{\alpha\mu}W^{\mu\beta}]\times [W_{\beta\nu}W^{\nu\alpha}],\\
\uppercase{o}_{T5}&=& Tr[W_{\mu\nu}W^{\mu\nu}]\times B_{\alpha\beta}B^{\alpha\beta},\\
\uppercase{o}_{T6}&=& Tr[W_{\alpha\nu}W^{\mu\beta}]\times B_{\beta\mu}B^{\alpha\nu},\\
\uppercase{o}_{T7}&=& Tr[W_{\alpha\mu}W^{\mu\beta}]\times B_{\beta\nu}B^{\nu\alpha},\\
\uppercase{o}_{T8}&=&[B_{\mu\nu}B^{\mu\nu}B_{\alpha\beta}B^{\alpha\beta}],\\
\uppercase{o}_{T9}&=& [B_{\alpha\nu}B^{\mu\beta}B_{\beta\nu}B^{\nu\alpha}].
\end{eqnarray}

All quartic gauge boson couplings altered with dimension-8 operators are presented in Table I.

Studies for the anomalous quartic $WW\gamma\gamma$ couplings have been carried out at the lepton colliders with the processes $e^{+}e^{-}\rightarrow VVV$ \cite{rodri,x1,x2,x3,x4,xf,ses,chen}, $e^{+}e^{-}\rightarrow VVFF$ \cite{x5,sen}, $e\gamma \rightarrow VVF$ \cite{x6,x7}, $\gamma\gamma \rightarrow VVV$ \cite{x9,x10}, $\gamma\gamma \rightarrow VV$ \cite{mur}, $e^{+}e^{-}\rightarrow e^{+}\gamma^{*}e^{-}\rightarrow VVFF$ \cite{sen1}, $e^{+}e^{-}\rightarrow e^{+}\gamma^{*}\gamma^{*}e^{-}\rightarrow e^{+}VVVe^{-}$ \cite{ff}, $e^{+}e^{-}\rightarrow e^{+}\gamma^{*}\gamma^{*}e^{-}\rightarrow e^{+} V V F$ \cite{sah} and at the hadron colliders with the processes $pp\rightarrow VVV$ \cite{x11,x12,x13,ahma,mar,wen,ye,dang}, $pp\rightarrow VVFF$ \cite{x14,per}, $p p \rightarrow p \gamma^{*}p \rightarrow p V V F$ \cite{banu} and $pp\rightarrow p\gamma^{*}\gamma^{*}p \rightarrow p V V p$ \cite{x15,x16,x17} where $V=W^{\pm},Z,\gamma$ and $F=e, j, \nu$.

The present experimental sensitivities on the anomalous $\frac{f_{M0}}{\Lambda^4}$, $\frac{f_{M1}}{\Lambda^4}$, $\frac{f_{M2}}{\Lambda^4}$ and $\frac{f_{M3}}{\Lambda^4}$ couplings arising from dimension-8 operators through the process $pp \rightarrow p\gamma^{*}\gamma^{*} p\rightarrow pW W p$ \cite{de1} at center-of-mass energy of $\sqrt{s}=8$ TeV using data corresponding to an integrated luminosity of 19.7 fb$^{-1}$ at the LHC are reported by CMS Collaboration. These are

\begin{eqnarray}
-4.2 &\textmd{TeV}^{-4}<\frac{f_{M0}}{\Lambda^4}<4.2 &\textmd{TeV}^{-4},
\end{eqnarray}
\begin{eqnarray}
-16 &\textmd{TeV}^{-4}<\frac{f_{M1}}{\Lambda^4}<16 &\textmd{TeV}^{-4},
\end{eqnarray}
\begin{eqnarray}
-2.1 &\textmd{TeV}^{-4}<\frac{f_{M2}}{\Lambda^4}<2.1 &\textmd{TeV}^{-4},
\end{eqnarray}
\begin{eqnarray}
-7.8 &\textmd{TeV}^{-4}<\frac{f_{M3}}{\Lambda^4}<7.8 &\textmd{TeV}^{-4}
\end{eqnarray}
at $95\%$ Confidence Level.

However, Ref. \cite{de} supplies the most restrictive limits on the anomalous quartic $\frac{f_{M4}}{\Lambda^4}$, $\frac{f_{M5}}{\Lambda^4}$, $\frac{f_{M6}}{\Lambda^4}$, $\frac{f_{M7}}{\Lambda^4}$, $\frac{f_{T0}}{\Lambda^4}$,$\frac{f_{T1}}{\Lambda^4}$, $\frac{f_{T2}}{\Lambda^4}$, $\frac{f_{T5}}{\Lambda^4}$, $\frac{f_{T6}}{\Lambda^4}$, $\frac{f_{T7}}{\Lambda^4}$ couplings which are related to the anomalous $WW\gamma\gamma$ quartic couplings derived with operators given by Eqs. 11-14 and 16-21. The results obtained for these couplings at $95\%$ Confidence Level through the process $pp \rightarrow W \gamma j j$ at $\sqrt{s}=8$ TeV with an integrated luminosity of 19.7 fb$^{-1}$ are listed as

\begin{eqnarray}
-40 &\textmd{TeV}^{-4}<\frac{f_{M4}}{\Lambda^4}<40 &\textmd{TeV}^{-4},
\end{eqnarray}
\begin{eqnarray}
-65 &\textmd{TeV}^{-4}<\frac{f_{M5}}{\Lambda^4}<65 &\textmd{TeV}^{-4},
\end{eqnarray}
\begin{eqnarray}
-129 &\textmd{TeV}^{-4}<\frac{f_{M6}}{\Lambda^4}<129 &\textmd{TeV}^{-4},
\end{eqnarray}
\begin{eqnarray}
-164 &\textmd{TeV}^{-4}<\frac{f_{M7}}{\Lambda^4}<162 &\textmd{TeV}^{-4},
\end{eqnarray}
\begin{eqnarray}
-5.4 &\textmd{TeV}^{-4}<\frac{f_{T0}}{\Lambda^4}<5.6 &\textmd{TeV}^{-4},
\end{eqnarray}
\begin{eqnarray}
-3.7 &\textmd{TeV}^{-4}<\frac{f_{T1}}{\Lambda^4}<4.0 &\textmd{TeV}^{-4},
\end{eqnarray}
\begin{eqnarray}
-11 &\textmd{TeV}^{-4}<\frac{f_{T2}}{\Lambda^4}<12 &\textmd{TeV}^{-4},
\end{eqnarray}
\begin{eqnarray}
-3.8 &\textmd{TeV}^{-4}<\frac{f_{T5}}{\Lambda^4}<3.8 &\textmd{TeV}^{-4},
\end{eqnarray}
\begin{eqnarray}
-2.8 &\textmd{TeV}^{-4}<\frac{f_{T6}}{\Lambda^4}<3.0 &\textmd{TeV}^{-4},
\end{eqnarray}
\begin{eqnarray}
-7.3 &\textmd{TeV}^{-4}<\frac{f_{T7}}{\Lambda^4}<7.7 &\textmd{TeV}^{-4}.
\end{eqnarray}

Recently, a lot of experimental and theoretical work have been done using dimension-6 operators for the anomalous quartic $WW\gamma\gamma$ couplings.
Dimension-6 operators can be determined in terms of dimension-8 operators with the simple relations. The relations between $f_{M_{i}}$ and $a_{0,c}$ couplings
are given as follows \cite{baa}

\begin{eqnarray}
\frac{f_{M0}}{\Lambda^4} =\frac{a_0}{\Lambda^2}\frac{1}{g^2v^2},
\end{eqnarray}

\begin{eqnarray}
\frac{f_{M1}}{\Lambda^4} =-\frac{a_c}{\Lambda^2}\frac{1}{g^2v^2},
\end{eqnarray}

\begin{eqnarray}
f_{M0}=f_{M,4}=\frac{f_{M2}}{2}=\frac{f_{M6}}{2},
\end{eqnarray}

\begin{eqnarray}
&f_{M,1}=\frac{f_{M3}}{2}=-\frac{f_{M5}}{2} =\frac{f_{M7}}{2}.
\end{eqnarray}

In this study, we examine 13 different anomalous $WW\gamma\gamma$ couplings at $95\%$ Confidence Level by considering diphoton production in the final state of the process $ep \rightarrow \nu_{e}\gamma \gamma j$ at 1.30, 1.98 TeV LHeC with an integrated luminosity of 100 fb$^{-1}$ and 7.07, 10 TeV FCC-he with an integrated luminosity of 1000 fb$^{-1}$. We obtain that our best limits are up to a factor of 10$^{2}$ better than the experimental limits. It can be seen from these results that the anomalous quartic $WW\gamma\gamma$ couplings can be examined with very good sensitivity at FCC-he.

\section{Numerical Analysis}

In our calculations, we analyze signals and backgrounds of the process $ep \rightarrow \nu_{e}\gamma\gamma j$ by using MadGraph$5_{-}$aMC$@$NLO \cite{mad} in which the anomalous quartic couplings are implemented through FeynRules package \cite{rul} through dimension-8 effective Lagrangians related to the anomalous quartic $WW\gamma\gamma$ couplings.  The CTEQ6L1 set is used to define the proton structure functions \cite{alp}. In order to obtain limits on the 13 different anomalous couplings arising from dimension-8 operators, we investigate the process $ep \rightarrow \nu_{e}\gamma \gamma j$. Symbolic diagram of this process is presented in Fig. 1.

In this study, we examine the background and signal processes in detail. First, the SM process with final state should be accepted as a background for the process $ep \rightarrow \nu_{e}\gamma\gamma j$. Also, we have considered the following background processes: $ep \rightarrow \nu_{e}\gamma\gamma \gamma j$ , $ep \rightarrow \nu_{e}\gamma\gamma j j$ , $ep \rightarrow \nu_{e} \nu_{\ell} \bar{\nu}_{\ell} \gamma \gamma j$. We know that the high dimensional operators could affect the $p_t$ distribution of photons, specially at the region with a large $p_t$ values, which can be very useful to distinguish signal and background events. Also, since two photons produced in the final state can come from the Higgs Boson, it can create a background for our process. For this reason, we apply a cut on the invariant mass of two photons. By applying a cut in the missing energy, we reduce the background to consider. Therefore, a set of cuts used for the analysis of signal and background processes are imposed as follows

\begin{eqnarray}
p_{T_{j}}>20 \, GeV, \quad p_{T_{\gamma}}>150 \, GeV, \quad p_{T_{\nu}}>20 GeV
\end{eqnarray}

\begin{eqnarray}
M_{\gamma \gamma}>200 GeV,
\end{eqnarray}

\begin{eqnarray}
 |\eta_{j}|< 5, \quad |\eta_{\gamma}|< 2.5,
\end{eqnarray}

\begin{eqnarray}
\Delta R(\gamma,\gamma)>0.4, \quad \Delta R(\gamma,j)>0.4,
\end{eqnarray}
where  $p_T$ is the transverse momentum of the final state particles, $\eta$ is the pseudorapidity and $\Delta R$ is the angular separation of the final state particles.

On the other hand, Fig.2(a) and Fig.2(b) show the effect of selected cuts (Eqs.42-45) on the anomalous quartic couplings. In the figures, the y-axis was labeled as cut efficiency. The cut efficiency is defined as follows; for each coupling values, the ratio of the cross section of selected cut set to the cross section of basic cut set(minimal cuts required to calculate cross sections). As seen in Fig.2(a) and Fig.2(b), the cut efficiency varies depend on the value of the anomalous quartic couplings and center-of-mass energy. The anomalous quartic $f_{ M_i}/\Lambda^4$ couplings are more sensitive than $f_{ T_i}/\Lambda^4$ couplings to these applied cut set. With the implementation of these cuts, the cross-section values of the processes involving $f_{ M_i}/\Lambda^4$ couplings decrease more than the processes involving $f_{ T_i}/\Lambda^4$ couplings. Despite these situations, background processes are more suppressed. Therefore, the resulting limits are better than the basic cut situation for $f_{ M_i}/\Lambda^4$ and $f_{ T_i}/\Lambda^4$ .

To have a comprehensive investigation on the cross section behavior, we present the analytical form of cross sections including the anomalous couplings,

\begin{eqnarray}
\sigma_{tot}(\frac{f_{i}}{\Lambda^{4}})=\sigma_{SM}+\frac{f_{i}}{\Lambda^{4}} \sigma_{int}+ \frac{f_{i}^{2}}{\Lambda^{8}} \sigma_{NP}  \ (i=1,...,13)
\end{eqnarray}
where $\sigma_{SM}$ shows the SM cross section, $\sigma_{int}$ and $\sigma_{NP}$ are the interference term between SM and the
new physics contribution, and the pure new physics contribution, respectively. In this analysis, we suppose that only one of the anomalous parameters deviates from the SM at any given time. We estimate the cross sections of the SM and signals after applying kinematic cuts used for the process $ep \rightarrow \nu_{e}\gamma \gamma j$ at the LHeC and FCC-he for 13 different anomalous couplings are given in Table II-III. As seen from Tables II-III, the largest deviation from the SM cross section takes place in $\frac{f_{T5}}{\Lambda^{4}}$ parameter among all anomalous couplings. For this reason, the limits on the anomalous $f_{T5}/\Lambda^{4}$ coupling are anticipated to be more sensitive  than the other anomalous couplings. A similar comment can be made between $f_{M2}/\Lambda^{4}$ coupling parameters and other $f_{Mi}/\Lambda^{4}$ ($i=0,1,3,4,5,7$) parameters.

The total cross sections of the process $ep \rightarrow \nu_{e}\gamma \gamma j$ as a function of 13 different anomalous couplings for the LHeC and the FCC-he are displayed in Figs. 3-15. As seen from these figures, the cross sections including new physics increase when the anomalous couplings grow in the interested range. Furthermore, we can see from these figures that the deviation from the SM of the cross sections including the anomalous couplings at center-of-mass energy of 10 TeV is larger than those of at center-of-mass energies of 1.30, 1.98, 7.07 TeV. Therefore, the obtained limits on new physics parameters at 10 TeV are expected to be more restrictive than the limits obtained from the other center-of-mass energies.

In order to investigate the limits on the anomalous quartic $WW\gamma\gamma$ couplings, we consider $\chi^{2}$ test with one-parameter sensitivity analysis. For this purpose, the $\chi^{2}$ test is described as follows

\begin{eqnarray}
\chi^{2}=\left(\frac{\sigma_{SM}-\sigma_{NP}}{\sigma_{SM}\delta}\right)^{2},
\end{eqnarray}
where $\delta=\frac{1}{\sqrt{\delta_{stat}^{2}}}, \delta_{stat}=\frac{1}{\sqrt{N_{SM}}}$ is the statistical error, $N_{SM}=L_{int}\times \sigma_{SM}$.

In Tables IV-VII, we present the sensitivities on the anomalous quartic $WW\gamma\gamma$ couplings for different center-of-mass energies and integrated luminosities. From Tables it is clear that increasing the integrated luminosity as well as center-of-mass energy provides more restricted limits on all the anomalous quartic couplings. Comparing the results in Table IV with the corresponding data in Table VII, there is an improvement in our limits up to several orders of magnitude with increasing integrated luminosity and center-of-mass energy. The best limits on these couplings are given for the FCC-he with 10 TeV in Table VII. As shown from Tables IV-V, since the LHeC has less center-of-mass energy and less luminosity than the LHC, sensitivity limits on the other anomalous quartic $WW\gamma\gamma$ couplings except for $f_{T5}/\Lambda^{4}$ obtained from our work are worse than the experimental limits. Our best limit obtained on $f_{T5}/\Lambda^{4}$ coupling is nearly 1.5 times better than the LHC. In Table VI, we present the sensitivity limits of $f_{T0}/\Lambda^{4}$, $f_{T1}/\Lambda^{4}$, $f_{T2}/\Lambda^{4}$, $f_{T5}/\Lambda^{4}$, $f_{T6}/\Lambda^{4}$ and $f_{T7}/\Lambda^{4}$ at $95\%$ Confidence Level through the process $ep \rightarrow \nu_{e}\gamma \gamma j$ at 7.07 TeV FCC-he. As can be seen from this Table, our best sensitivities on these couplings are up to one order of magnitude better than the sensitivities derived in Ref. \cite{de1}. The most important results on $f_{M0}/\Lambda^{4}$, $f_{M1}/\Lambda^{4}$, $f_{M2}/\Lambda^{4}$ and $f_{M3}/\Lambda^{4}$ couplings given in Table VII are comparable to the limits obtained from Ref. \cite{de1}. The best limit in $f_{Mi}/\Lambda^{4}$ ($i=0,1,2,3,4,5,7$) parameters is obtained for $f_{M2}/\Lambda^{4}$ parameter. For FCC-he with a center-of-mass energy of 10 TeV and an integrated luminosity 1000 fb$^{-1}$, the sensitivity on $f_{M2}/\Lambda^{4}$ coupling are found as [-0.54;0.47] (TeV$^{-4}$). The estimated sensitivities of the FCC-he for $f_{T1}/\Lambda^{4}$, $f_{T2}/\Lambda^{4}$ and $f_{T7}/\Lambda^{4}$ couplings at $95\%$ Confidence Level are at the order of $10^{-1}$ TeV$^{-4}$. In addition, it can be understood from Table VII that the limits on $f_{M4}/\Lambda^{4}$, $f_{M5}/\Lambda^{4}$ and $f_{M7}/\Lambda^{4}$ presented by Ref. \cite{de} are worse than the sensitivities at 100 fb$^{-1}$. Our limits on $f_{T0}/\Lambda^{4}$, $f_{T1}/\Lambda^{4}$, $f_{T2}/\Lambda^{4}$ and $f_{T7}/\Lambda^{4}$ couplings are roughly one order better than those with respect to the best sensitivity derived from the LHC. Finally, the limits on $f_{T5}/\Lambda^{4}$ and $f_{T6}/\Lambda^{4}$ couplings at $\sqrt{s}=10$ TeV and $L_{int}=1000$ fb$^{-1}$ is [-2.42;2.26] $\times 10^{-2}$ TeV$^{-4}$ and [-5.91;5.54] $\times 10^{-2}$ TeV$^{-4}$.

\section{Conclusions}

The $ep$ colliders such as the LHeC and the FCC-he would significantly enrich the physics reachable with the LHC. These colliders may provide a lot of important information to test new physics effects beyond the SM and the measurements of the SM. In this work, we offer a study to constrain new physics with the anomalous quartic $WW\gamma\gamma$ gauge boson couplings defined by effective Lagrangian method. In the literature, the new physics effects of quartic gauge boson couplings are usually examined in a model independent way by means of the effective Lagrangian approach. These couplings are described by dimension-8 operators that have very strong energy dependency with respect to the SM. Therefore, the total cross sections containing the anomalous quartic couplings are expected to be greater than the cross sections of the SM.  In this case, any possible deviation from the SM predictions of the examined process would be a sign for the presence of new physics beyond the SM.

As far as we can see from the literature, this study is the first report on the anomalous quartic couplings determined by effective Lagrangians at $ep$ colliders. Moreover, we consider that this paper will motivate further works to investigate the another anomalous quartic couplings at $ep$ colliders.

Consequently, the process $ep \rightarrow \nu_{e}\gamma \gamma j$ is very beneficial to sensitivity studying on the anomalous quartic $WW\gamma\gamma$ couplings and illustrates the complementarity between LHC and future $ep$ colliders for probing extensions of the SM.

\textbf{Acknowledgement}:
The numerical calculations reported in this paper were fully performed at TUBITAK ULAKBIM, High Performance and Grid Computing Center (TRUBA resources).

\pagebreak

\pagebreak

\begin{figure}
\includegraphics [width=7cm,height=4cm]{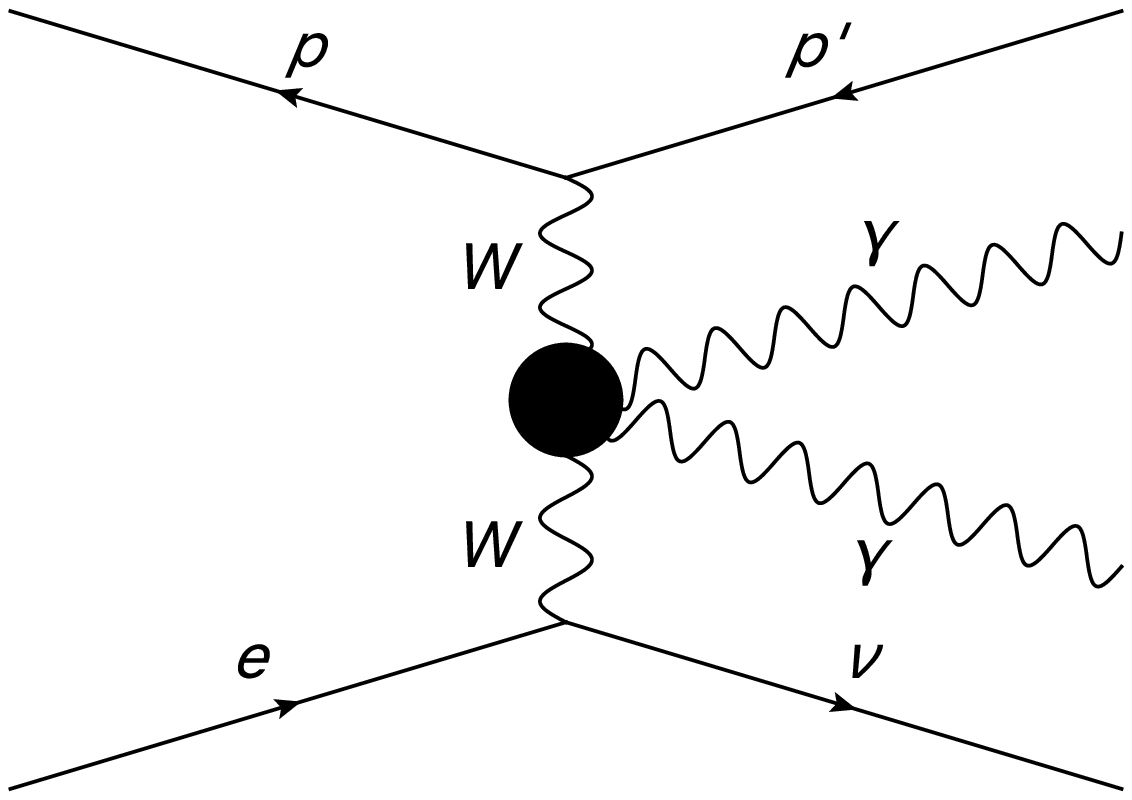}
\caption{Diagram for the process $ep \rightarrow \nu_{e}\gamma\gamma j$.
\label{fig2}}
\end{figure}

\begin{figure}
    \centering
    \subfloat[]{{\includegraphics[width=7.7cm]{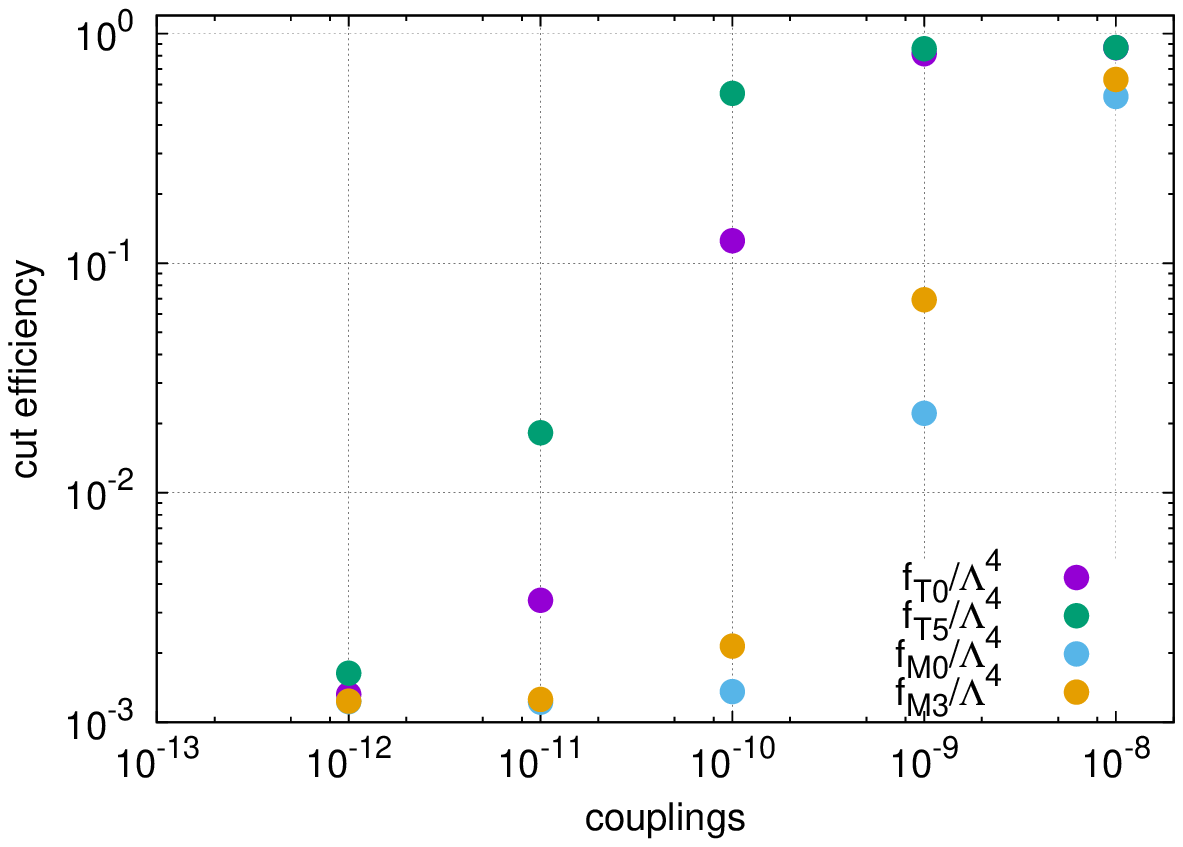} }}
    \qquad
    \subfloat[]{{\includegraphics[width=7.7cm]{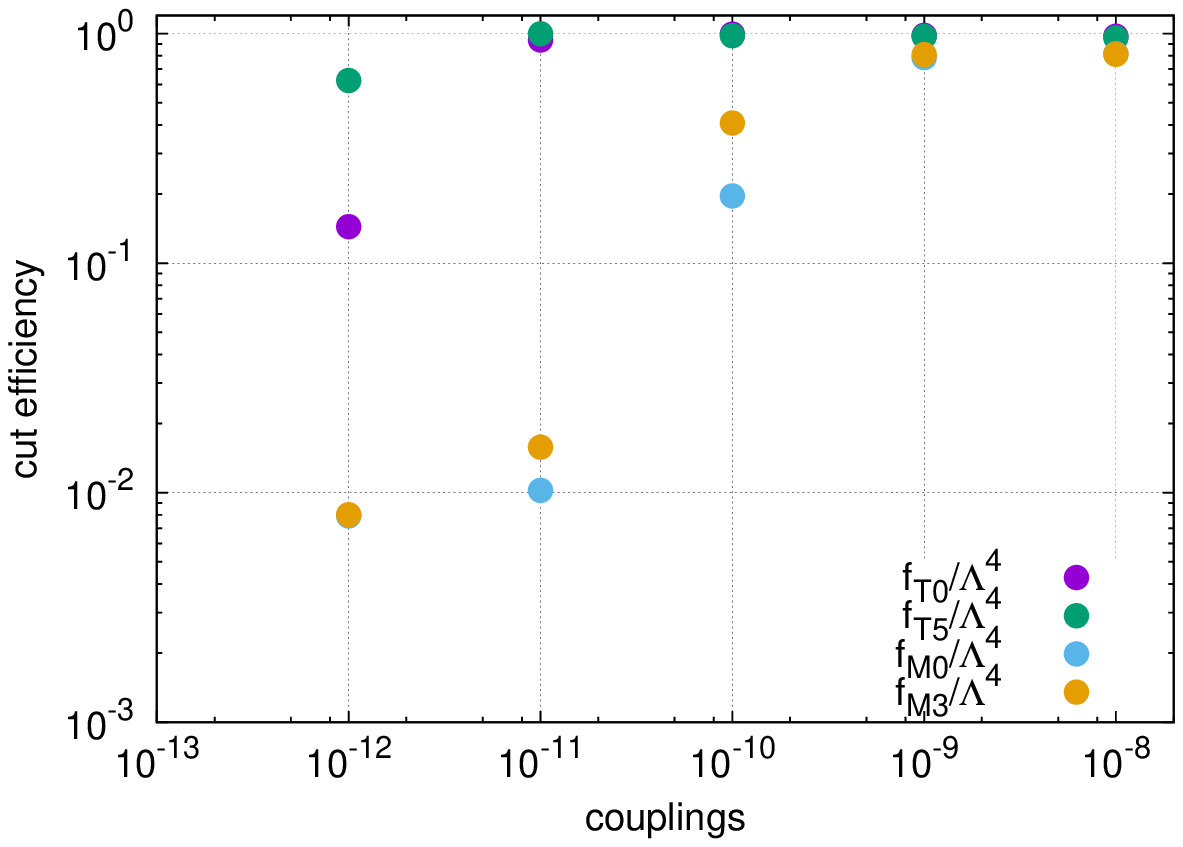} }}
    \caption{Fig.2(a) and Fig.2(b) show the variation of cut efficiency for some anomalous quartic parameters with respect to different values of these couplings at 1.98 TeV and 10 TeV center-of-mass energies, respectively.}
    \label{fig:example}
\end{figure}

\begin{figure}
\includegraphics{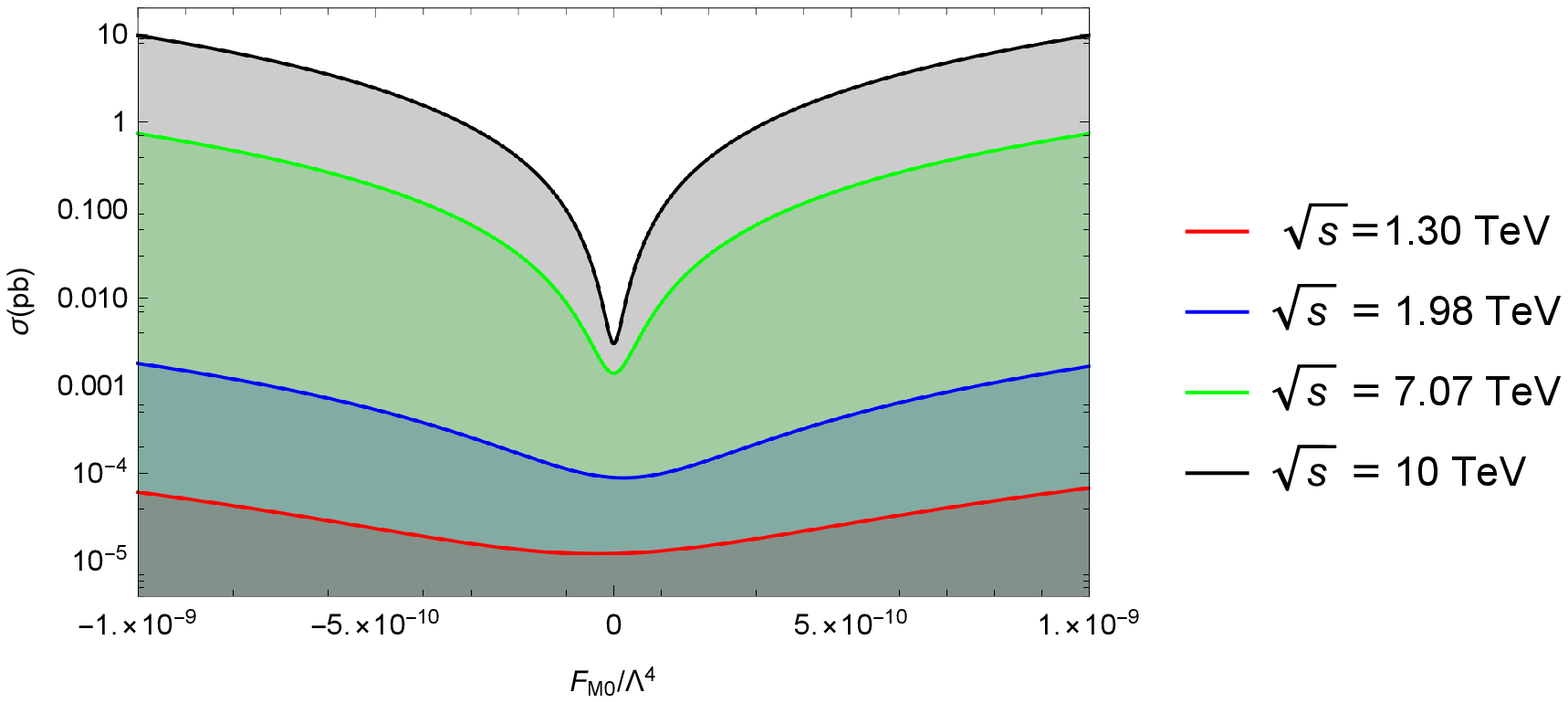}
\caption{The total cross sections of the process $ep \rightarrow \nu_{e}\gamma\gamma j$ depending on the anomalous $\frac{f_{M0}}{\Lambda^4}$ coupling at the LHeC and FCC-he.
\label{fig2}}
\end{figure}

\begin{figure}
\includegraphics{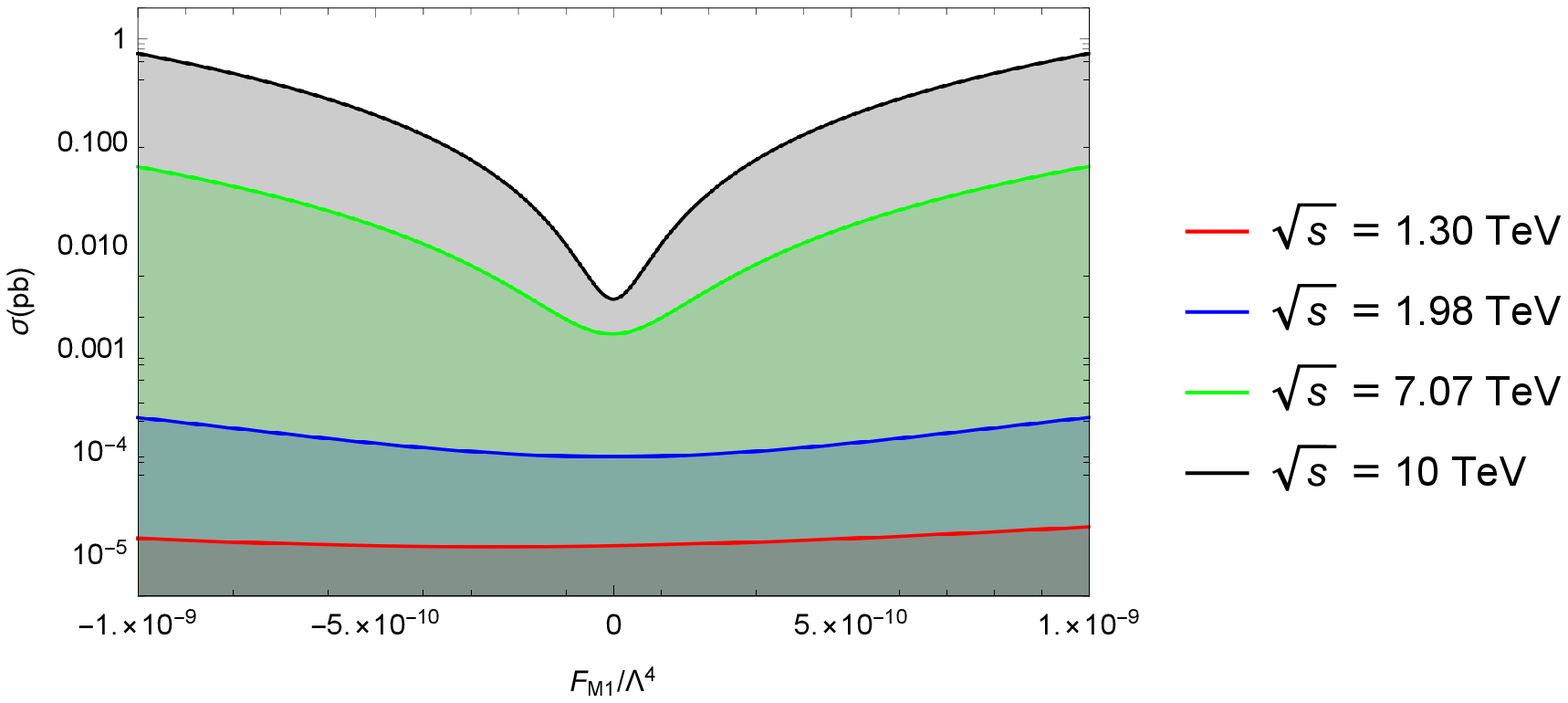}
\caption{The same as Fig. 2 but for $\frac{f_{M1}}{\Lambda^4}$.
\label{fig3}}
\end{figure}

\begin{figure}
\includegraphics{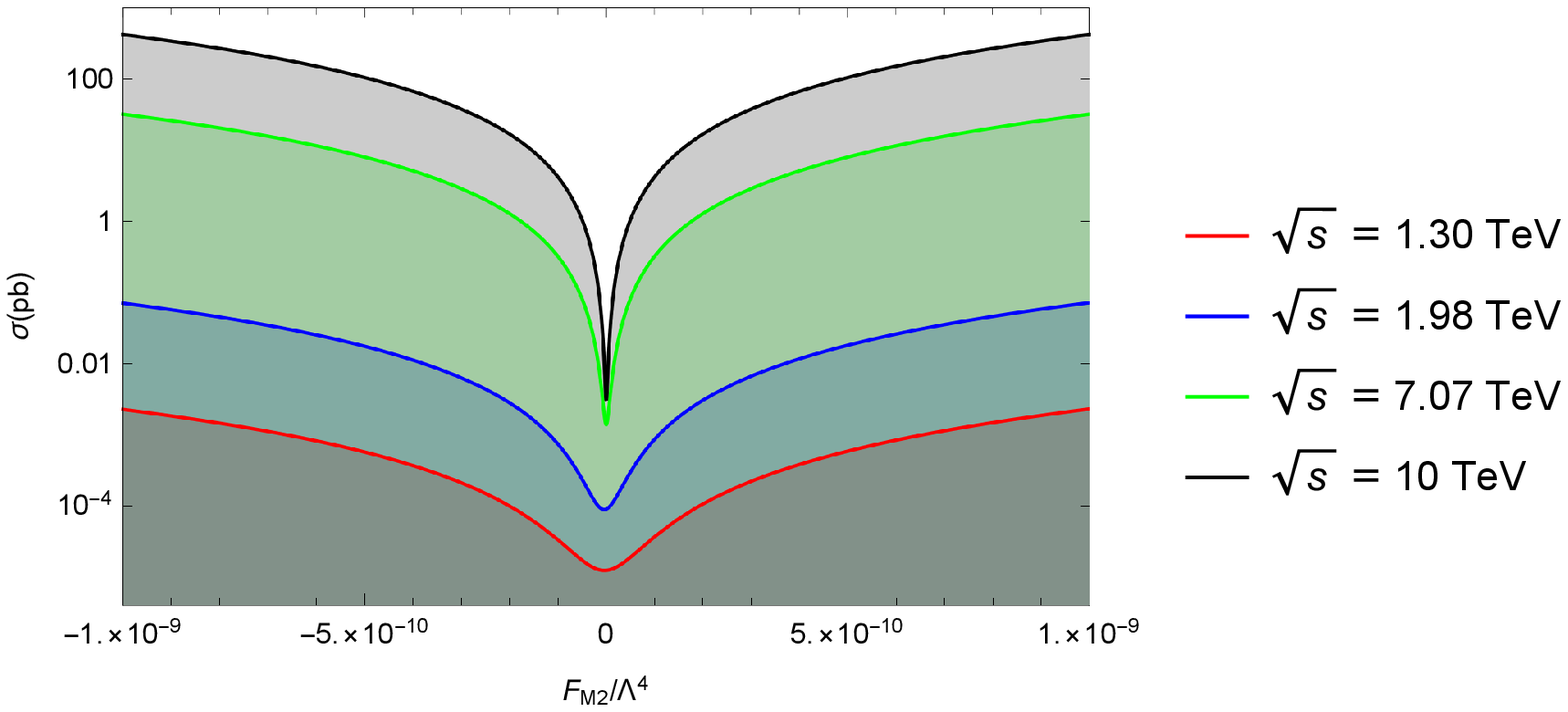}
\caption{The same as Fig. 2 but for $\frac{f_{M2}}{\Lambda^4}$.
\label{fig4}}
\end{figure}

\begin{figure}
\includegraphics{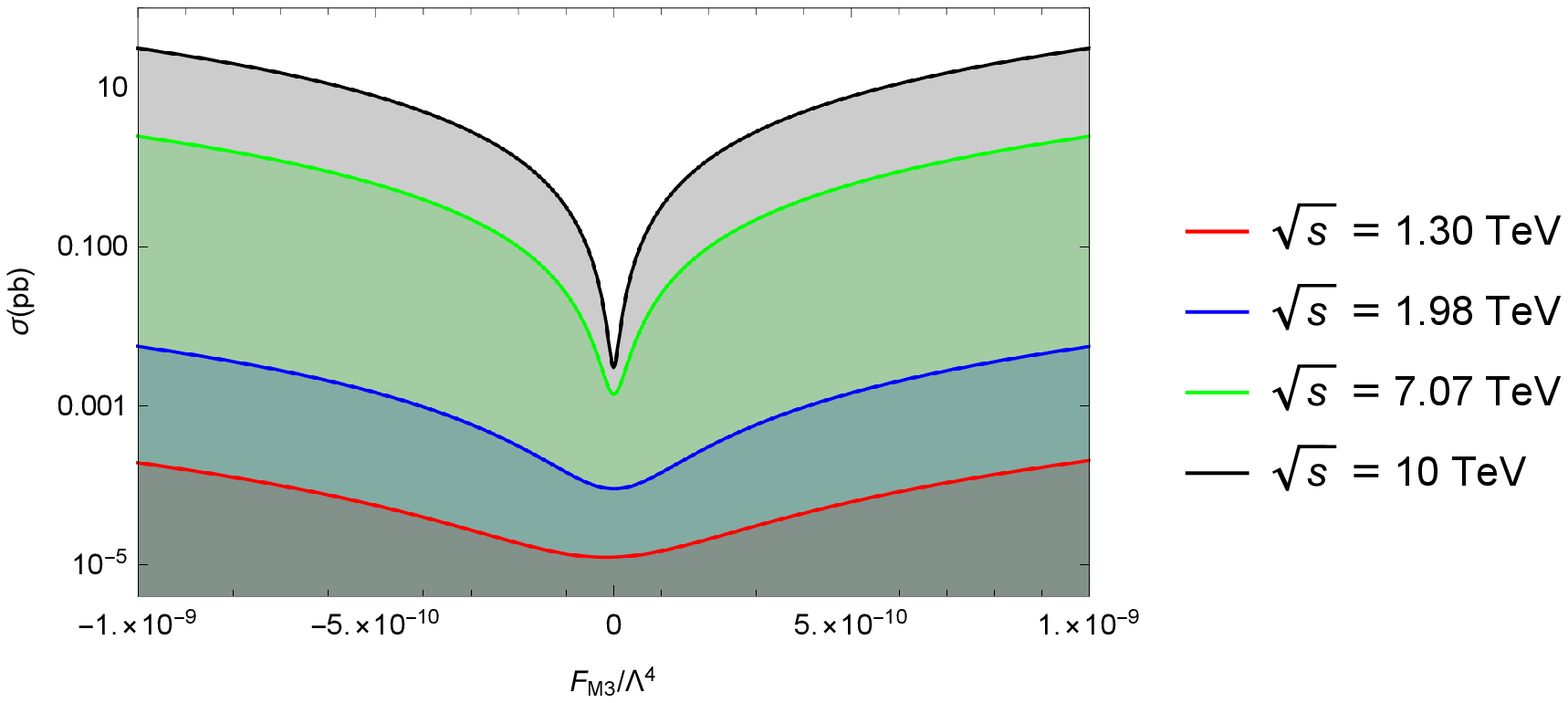}
\caption{The same as Fig. 2 but for $\frac{f_{M3}}{\Lambda^4}$.
\label{fig5}}
\end{figure}

\begin{figure}
\includegraphics{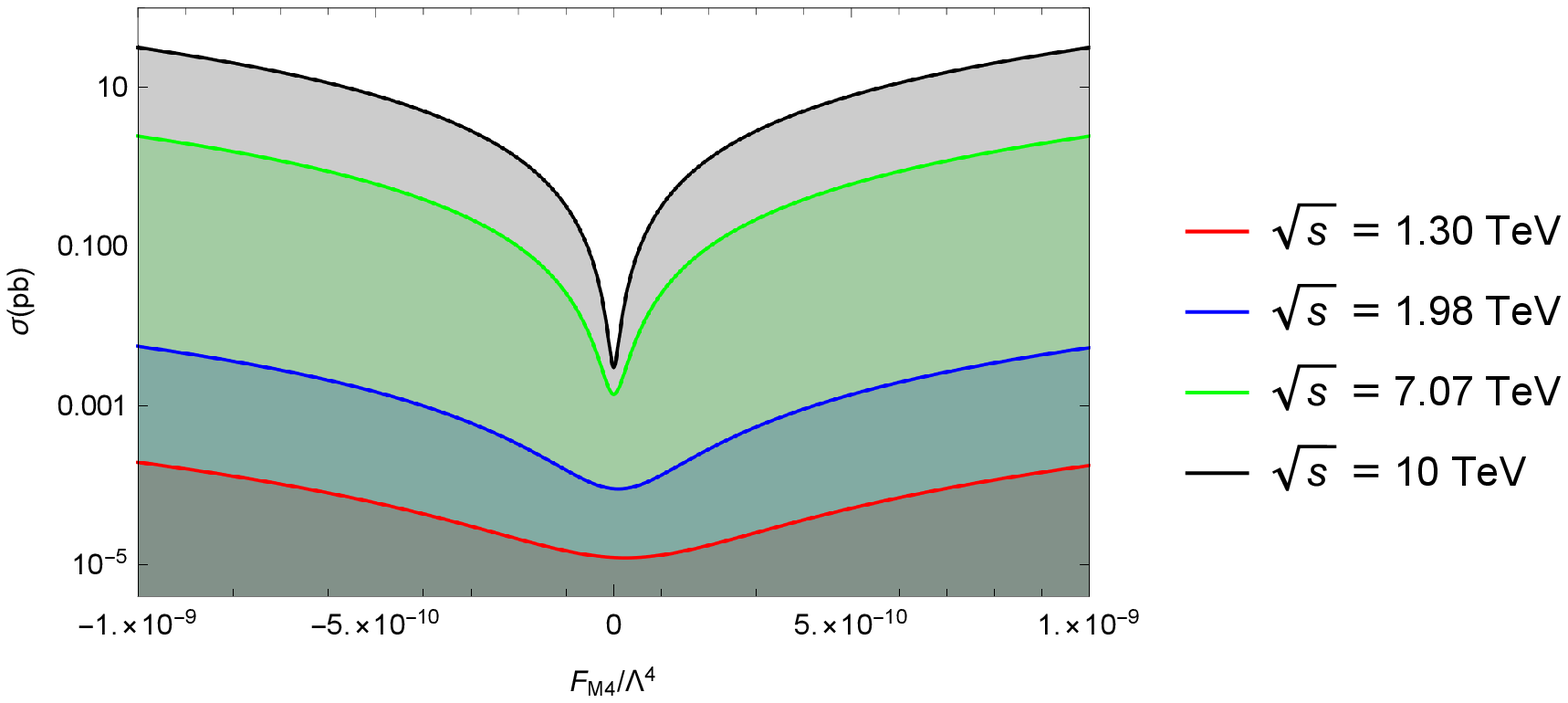}
\caption{The same as Fig. 2 but for $\frac{f_{M4}}{\Lambda^4}$.
\label{fig6}}
\end{figure}

\begin{figure}
\includegraphics{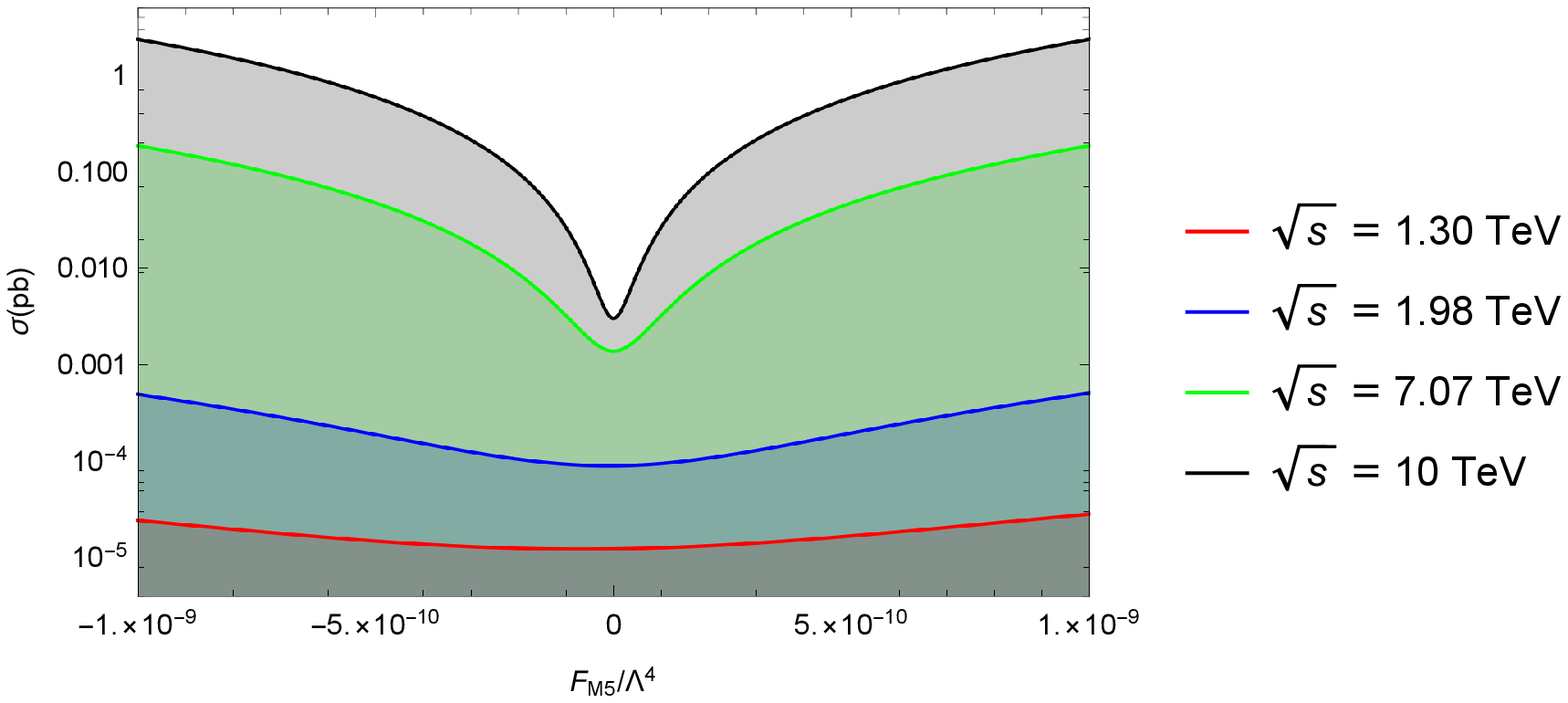}
\caption{The same as Fig. 2 but for $\frac{f_{M5}}{\Lambda^4}$.
\label{fig7}}
\end{figure}

\begin{figure}
\includegraphics{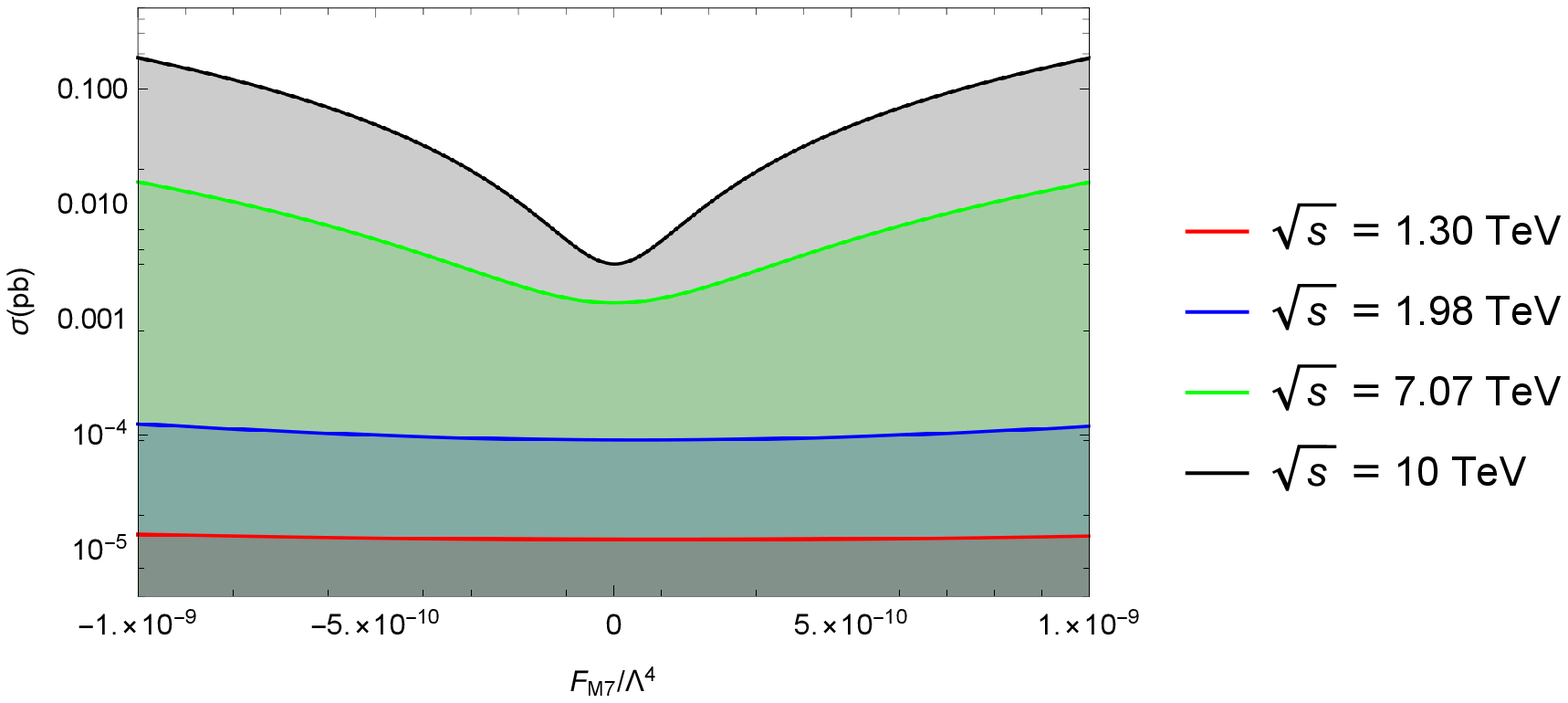}
\caption{The same as Fig. 2 but for $\frac{f_{M7}}{\Lambda^4}$.
\label{fig8}}
\end{figure}

\begin{figure}
\includegraphics{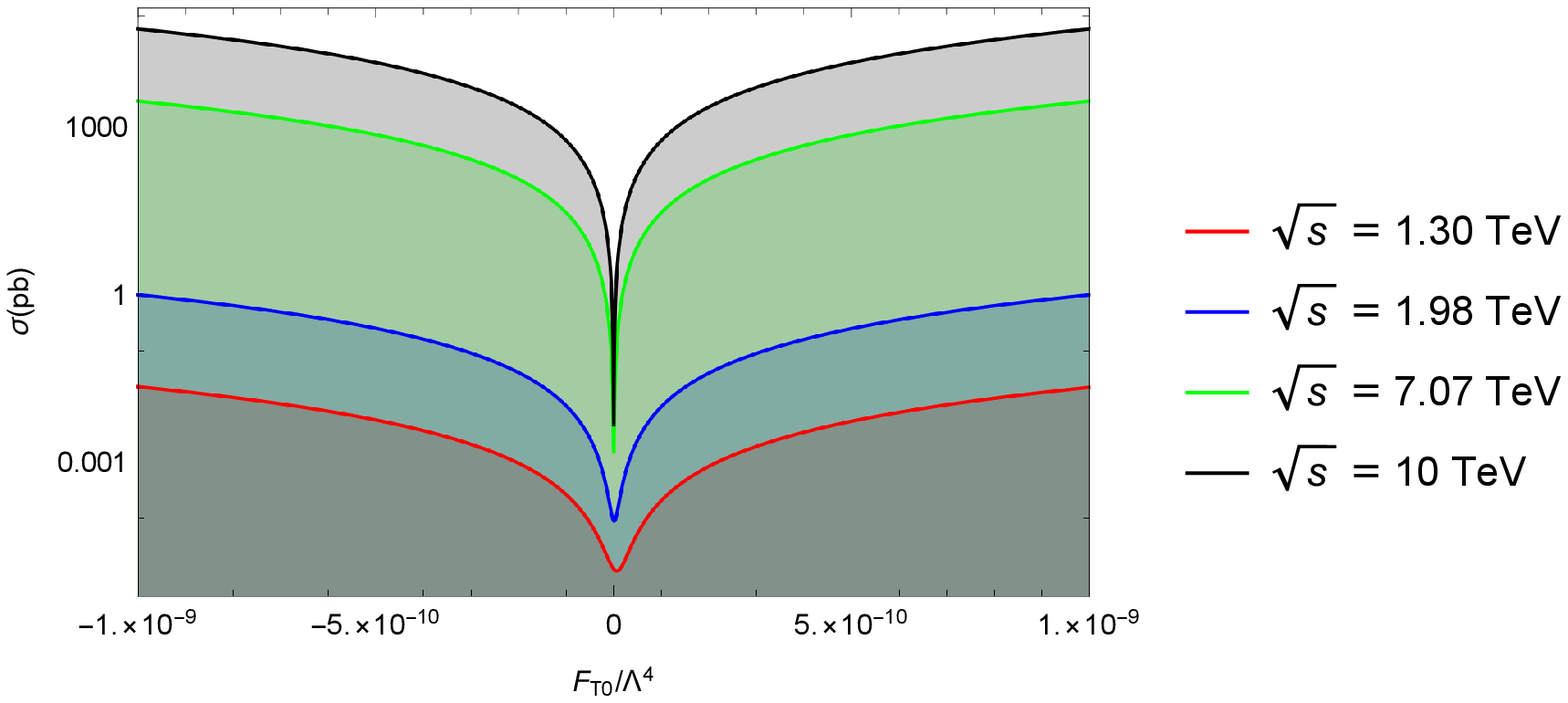}
\caption{The total cross sections of the process $ep \rightarrow \nu_{e}\gamma\gamma j$ depending on the anomalous $\frac{f_{T0}}{\Lambda^4}$ coupling at the LHeC and FCC-he.
\label{fig9}}
\end{figure}

\begin{figure}
\includegraphics{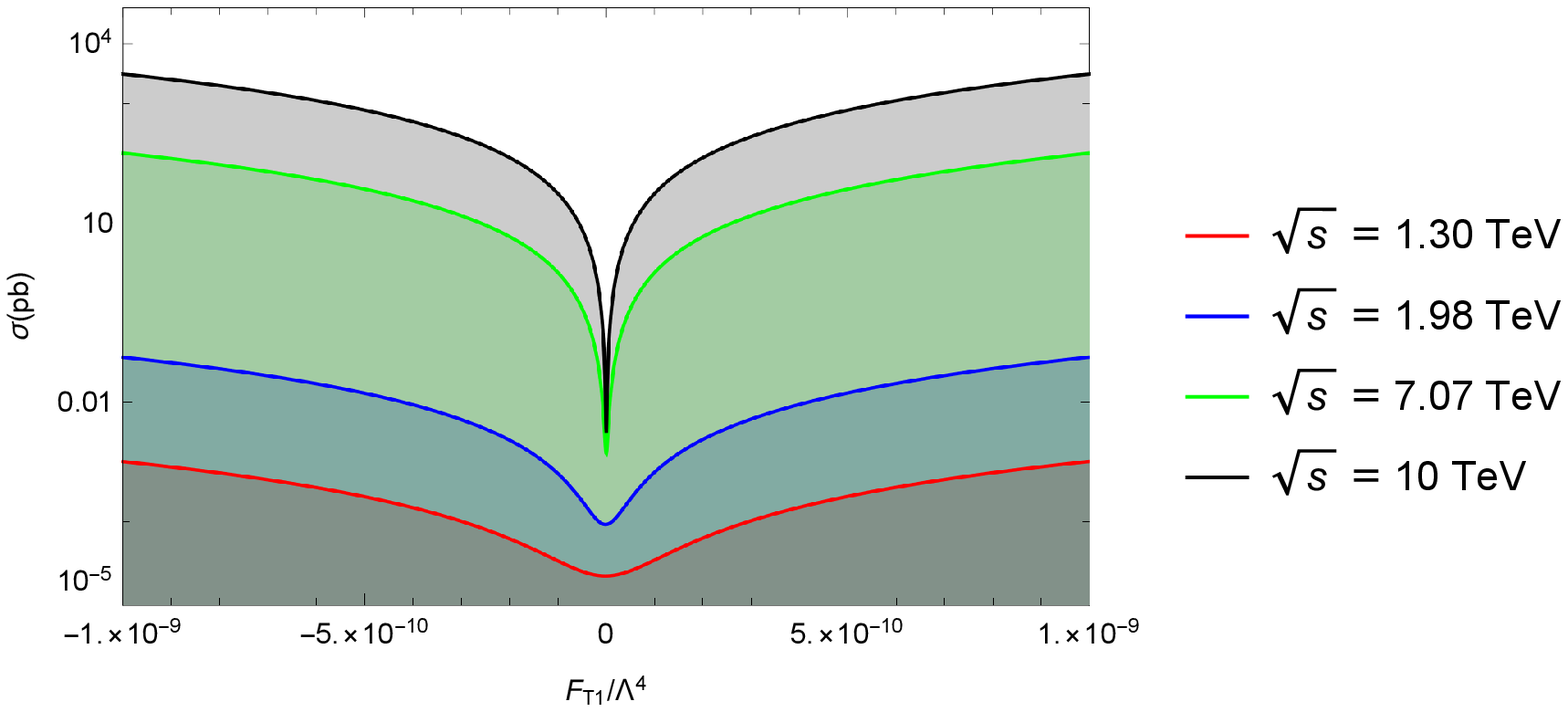}
\caption{The same as Fig. 9 but for $\frac{f_{T1}}{\Lambda^4}$.
\label{fig10}}
\end{figure}

\begin{figure}
\includegraphics{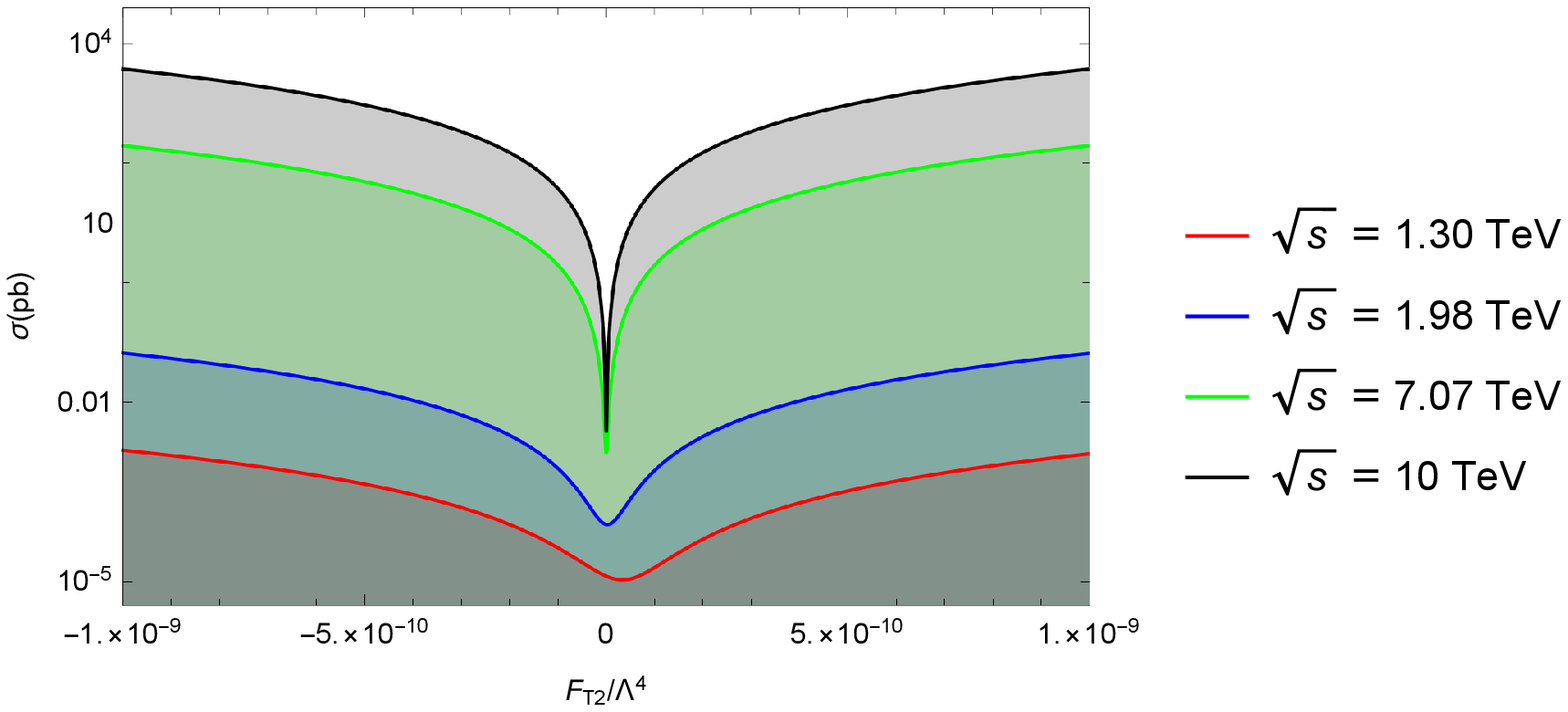}
\caption{The same as Fig. 9 but for $\frac{f_{T2}}{\Lambda^4}$.
\label{fig11}}
\end{figure}

\begin{figure}
\includegraphics{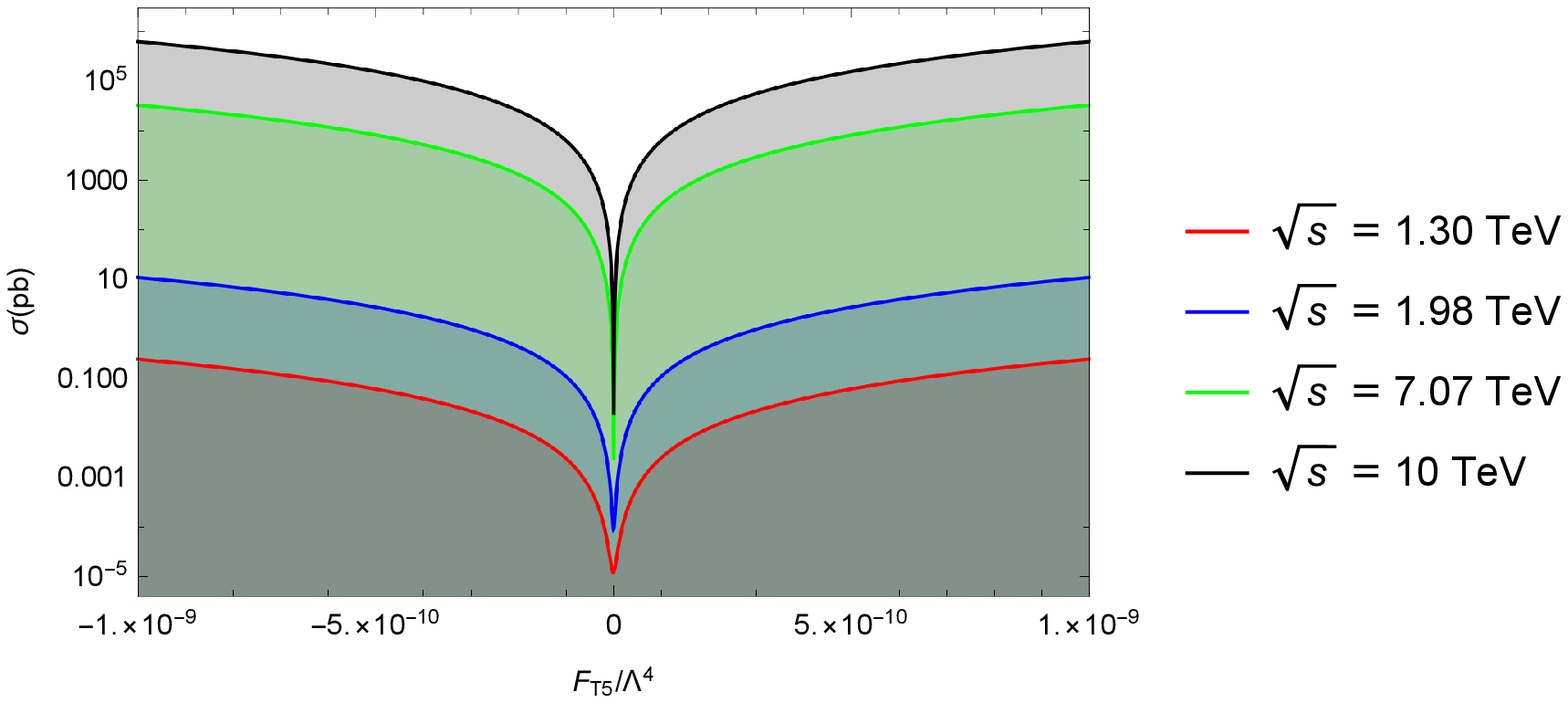}
\caption{The same as Fig. 9 but for $\frac{f_{T5}}{\Lambda^4}$.
\label{fig12}}
\end{figure}

\begin{figure}
\includegraphics{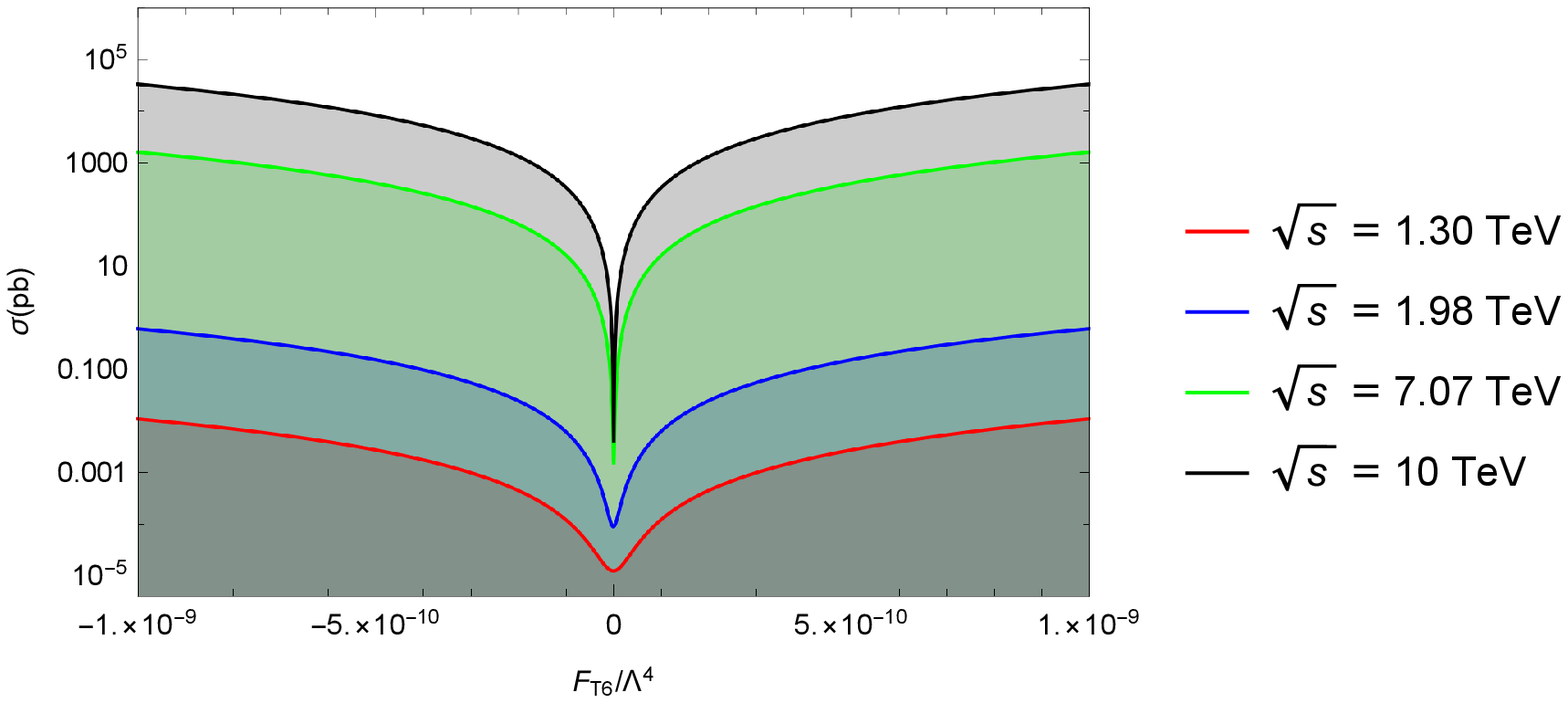}
\caption{The same as Fig. 9 but for $\frac{f_{T6}}{\Lambda^4}$.
\label{fig13}}
\end{figure}

\begin{figure}
\includegraphics{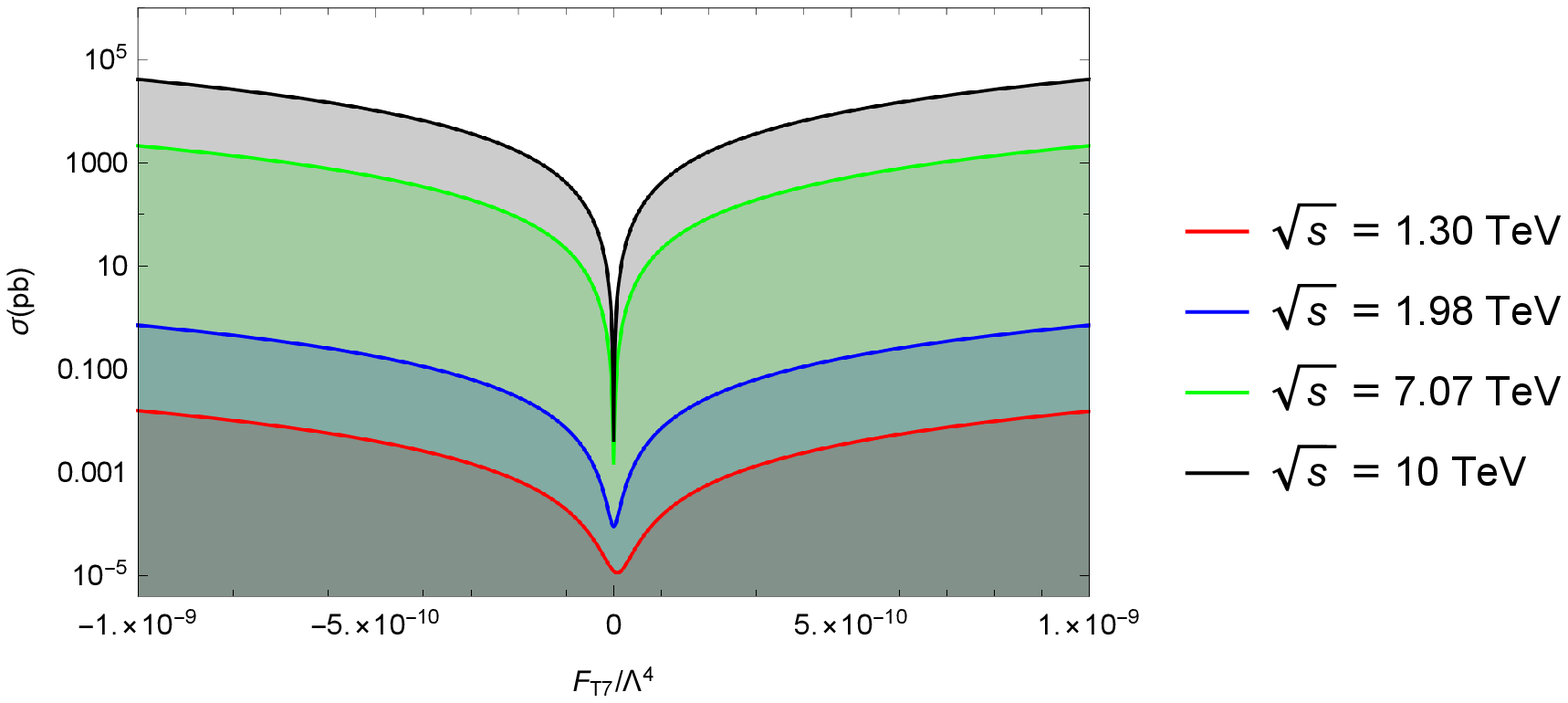}
\caption{The same as Fig. 9 but for $\frac{f_{T7}}{\Lambda^4}$.
\label{fig14}}
\end{figure}

\begin{table}
\caption{The quartic gauge boson couplings altered with dimension-8 operators are shown with X.}
\begin{ruledtabular}
\begin{tabular}{ccccccccccccc}
&&$WWWW$ &$WWZZ$ & $ZZZZ$ & $WW\gamma Z$ & $WW\gamma \gamma$ & $ZZZ\gamma$ & $ZZ \gamma \gamma$ & $Z\gamma \gamma \gamma$ & $\gamma\gamma\gamma\gamma$ &\\
\hline
$\uppercase{o}_{S0},\uppercase{o}_{S1}$ &&$X$  &$X$ &$X$ & \\
\hline
$\uppercase{o}_{M0},\uppercase{o}_{M1},\uppercase{o}_{M6},\uppercase{o}_{M7}$ &&$X$ &$X$ &$X$ &$X$ &$X$ &$X$ &$X$  &\\
\hline
$\uppercase{o}_{M2},\uppercase{o}_{M3},\uppercase{o}_{M4},\uppercase{o}_{M5}$&&&$X$ &$X$ &$X$ &$X$ &$X$ &$X$ & \\
\hline
$\uppercase{o}_{T0},\uppercase{o}_{T1},\uppercase{o}_{T2}$&&$X$ &$X$ &$X$ &$X$ &$X$ &$X$ &$X$ &$X$ &$X$ & \\
\hline
$\uppercase{o}_{T5},\uppercase{o}_{T6},\uppercase{o}_{T7}$&&&$X$ &$X$ &$X$ &$X$ &$X$ &$X$ &$X$ &$X$ &\\
\hline
$\uppercase{o}_{T8},\uppercase{o}_{T9}$ &&&&$X$&& &$X$ &$X$ &$X$ &$X$ & \\
\hline
\end{tabular}
\end{ruledtabular}
\end{table}

\begin{table}
\caption{The cross sections of signal for $10^{3}$ TeV$^{-4}$ value of anomalous $f_{M0}/\Lambda^{4}$, $f_{M1}/\Lambda^{4}$,$f_{M2}/\Lambda^{4}$, $f_{M3}/\Lambda^{4}$, $f_{M4}/\Lambda^{4}$, $f_{M5}/\Lambda^{4}$, $f_{M7}/\Lambda^{4}$, $f_{T0}/\Lambda^{4}$, $f_{T1}/\Lambda^{4}$, $f_{T2}/\Lambda^{4}$, $f_{T5}/\Lambda^{4}$, $f_{T6}/\Lambda^{4}$ and $f_{T7}/\Lambda^{4}$ after cuts given in Eqs. 45-48 of the process $ep \rightarrow \nu_{e}W^{+} W^{-}j\rightarrow \nu_{e} \ell^{+}\nu_{\ell}\ell^{-}\nu_{\ell}j$ at the LHeC with $\sqrt{s}=1.30$ and $1.98$ TeV.
All background cross sections of the process for $\sqrt{s}=1.30$ and $1.98$ TeV are $1.23\times 10^{-5}$ pb and $9.01\times 10^{-5}$ pb, respectively.}
\begin{tabular}{|c|c|c|}\hline
Couplings (TeV$^{-4}$)  & $\sigma_{ep} (\mbox{pb})@ \sqrt{s}=1.30$ TeV &  $\sigma_{ep}(\mbox{pb})@ \sqrt{s}=1.98$ TeV    \\
\hline \hline
$f_{M0}/\Lambda^{4}$  & 5.79 $\times 10^{-5}$ & 1.64  $\times 10^{-3}$ \\
\hline
$f_{M1}/\Lambda^{4}$  & 1.88 $\times 10^{-5}$ & 2.38 $ \times 10^{-4}$ \\
\hline
$f_{M2}/\Lambda^{4}$  & 2.23  $\times 10^{-3}$ & 7.16 $\times 10^{-2}$  \\
\hline
$f_{M3}/\Lambda^{4}$  & 2.10 $\times 10^{-4}$  & 5.64 $\times 10^{-3}$ \\
\hline
$f_{M4}/\Lambda^{4}$  & 1.74 $\times 10^{-4}$  & 5.39 $\times 10^{-3}$ \\
\hline
$f_{M5}/\Lambda^{4}$  & 2.26 $\times 10^{-5}$  & 4.75 $\times 10^{-4}$ \\
\hline
$f_{M7}/\Lambda^{4}$  & 1.25 $\times 10^{-5}$  & 1.12 $\times 10^{-4}$ \\
\hline
$f_{T0}/\Lambda^{4}$  &  2.37  $\times 10^{-2}$ & 1.01   \\
\hline
$f_{T1}/\Lambda^{4}$  &  1.07 $\times 10^{-3}$ & 5.75  $\times 10^{-2}$  \\
\hline
$f_{T2}/\Lambda^{4}$  &  1.64 $\times 10^{-3}$ & 6.79  $\times 10^{-2}$  \\
\hline
$f_{T5}/\Lambda^{4}$  & 2.50 $\times 10^{-1}$  & 1.08 $\times 10^{1}$  \\
\hline
$f_{T6}/\Lambda^{4}$  & 1.11  $\times 10^{-2}$ & 6.16 $\times 10^{-1}$  \\
\hline
$f_{T7}/\Lambda^{4}$  & 1.65  $\times 10^{-2}$ & 7.13 $\times 10^{-1}$  \\
\hline
\end{tabular}
\end{table}

\begin{table}
\caption{The same as Table II but for $\sqrt{s}=7.07$ and $10$ TeV. All background cross sections of the process for $\sqrt{s}=7.07$ and $10$ TeV are $1.39\times 10^{-3}$ pb and $3.03\times 10^{-3}$ pb, respectively.}
\begin{tabular}{|c|c|c|}\hline
\hline\hline
Couplings (TeV$^{-4}$)  & $\sigma_{ep} (\mbox{pb})@ \sqrt{s}=7.07$ TeV &  $\sigma_{ep}(\mbox{pb})@ \sqrt{s}=10$ TeV    \\
\hline \hline
$f_{M0}/\Lambda^{4}$  & 8.67 $\times 10^{-3}$ & 9.98  $\times 10^{-2}$ \\
\hline
$f_{M1}/\Lambda^{4}$  & 1.97 $\times 10^{-3}$ & 1.04 $ \times 10^{-2}$ \\
\hline
$f_{M2}/\Lambda^{4}$  & 3.19  $\times 10^{-1}$ & 4.24  \\
\hline
$f_{M3}/\Lambda^{4}$  & 2.56 $\times 10^{-2}$  & 3.23 $\times 10^{-1}$ \\
\hline
$f_{M4}/\Lambda^{4}$  & 2.52 $\times 10^{-2}$  & 3.27 $\times 10^{-1}$ \\
\hline
$f_{M5}/\Lambda^{4}$  & 3.16 $\times 10^{-3}$  & 2.68 $\times 10^{-2}$ \\
\hline
$f_{M7}/\Lambda^{4}$  & 1.52 $\times 10^{-3}$  & 4.81 $\times 10^{-3}$ \\
\hline
$f_{T0}/\Lambda^{4}$  &  3.06  $\times 10^{1}$  & 5.87 $\times 10^{2}$   \\
\hline
$f_{T1}/\Lambda^{4}$  &  1.47 & 3.06  $\times 10^{1}$  \\
\hline
$f_{T2}/\Lambda^{4}$  &  2.01 & 3.75  $\times 10^{1}$  \\
\hline
$f_{T5}/\Lambda^{4}$  & 3.29 $\times 10^{2}$  & 6.33 $\times 10^{3}$  \\
\hline
$f_{T6}/\Lambda^{4}$  & 1.61  $\times 10^{1}$  & 3.34 $\times 10^{2}$  \\
\hline
$f_{T7}/\Lambda^{4}$  & 2.15  $\times 10^{1}$ & 4.01 $\times 10^{2}$  \\
\hline
\end{tabular}
\end{table}

\begin{table}
\caption{The limits obtained on the anomalous quartic $WW\gamma\gamma$ couplings at $95\%$ Confidence Level through the process $ep \rightarrow \nu_{e}\gamma \gamma j$ at 1.30 TeV LHeC with integrated luminosities of 10, 30, 50, 100 fb$^{-1}$.}
\begin{tabular}{|c|c|c|c|c|c|}\hline
Couplings (TeV$^{-4}$) & 10 fb$^{-1}$ & 30 fb$^{-1}$ & 50 fb$^{-1}$ & 100 fb$^{-1}$ \\
\hline \hline
$f_{M0}/\Lambda^{4}$  & [-1.18;1.11] $\times 10^{3}$ & [-0.91;0.84] $\times 10^{3}$ & [-0.80;0.73] $\times 10^{3}$ & [-0.68;0.61] $\times 10^{3}$ \\
\hline
$f_{M1}/\Lambda^{4}$  & [-4.27;3.77] $\times 10^{3}$ & [-3.31;2.80]$ \times 10^{3}$  & [-2.95;2.45] $ \times 10^{3}$ & [-2.52;2.02] $ \times 10^{3}$   \\
\hline
$f_{M2}/\Lambda^{4}$  & [-1.79;1.71] $\times 10^{2}$ & [-1.37;1.29]$ \times 10^{2}$  & [-1.21;1.13] $ \times 10^{2}$ & [-1.02;0.95]$\times 10^{2}$   \\
\hline
$f_{M3}/\Lambda^{4}$  & [-6.29;5.96] $\times 10^{2}$ & [-4.82;4.49]$ \times 10^{2}$  & [-4.27;3.94] $ \times 10^{2}$ & [-3.62;3.28]$\times 10^{2}$   \\
\hline
$f_{M4}/\Lambda^{4}$  & [-6.08;6.57] $\times 10^{2}$ & [-4.57;5.06]$ \times 10^{2}$  & [-4.00;4.48] $ \times 10^{2}$ & [-3.32;3.81]$\times 10^{2}$   \\
\hline
$f_{M5}/\Lambda^{4}$  & [-2.30;2.16] $\times 10^{3}$ & [-1.76;1.62]$ \times 10^{3}$  & [-1.56;1.42] $ \times 10^{3}$ & [-1.32;1.18]$\times 10^{3}$   \\
\hline
$f_{M7}/\Lambda^{4}$  & [-0.80;0.82]$\times 10^{4}$    & [-0.60;0.62]$\times 10^{4}$   & [-0.53;0.55]$\times 10^{4}$   & [-0.44;0.46]$\times 10^{4}$    \\
\hline
$f_{T0}/\Lambda^{4}$  &  [-0.49;0.62] $\times 10^{2}$ & [-0.36;0.49] $\times 10^{2}$ & [-0.31;0.44] $\times 10^{2}$ & [-0.25;0.38] $\times 10^{2}$ \\
\hline
$f_{T1}/\Lambda^{4}$  & [-2.64;2.60] $\times 10^{2}$  & [-2.01;1.97] $\times 10^{2}$ & [-1.77;1.74] $\times 10^{2}$ & [-1.49;1.46] $\times 10^{2}$  \\
\hline
$f_{T2}/\Lambda^{4}$  &  [-1.87;2.54] $\times 10^{2}$ & [-1.35;2.02] $\times 10^{2}$ & [-1.16;1.82] $\times 10^{2}$ & [-0.93;1.60] $\times 10^{2}$ \\
\hline
$f_{T5}/\Lambda^{4}$  & [-1.78;1.59] $\times 10^{1}$  & [-1.38;1.19] $\times 10^{1}$ & [-1.23;1.03] $\times 10^{1}$ & [-1.05;0.86] $\times 10^{1}$  \\
\hline
$f_{T6}/\Lambda^{4}$  & [-8.04;7.92] $\times 10^{1}$  & [-6.12;6.01] $\times 10^{1}$ & [-5.40;5.28] $\times 10^{1}$ & [-4.55;4.43] $\times 10^{1}$ \\
\hline
$f_{T7}/\Lambda^{4}$  & [-0.60;0.75] $\times 10^{2}$  & [-0.44;0.59] $\times 10^{2}$ & [-0.38;0.53] $\times 10^{2}$ & [-0.31;0.46] $\times 10^{2}$ \\
\hline
\end{tabular}
\end{table}

\begin{table}
\caption{The same as Table IV but for 1.98 TeV LHeC.}
\begin{tabular}{|c|c|c|c|c|c|}\hline
Couplings (TeV$^{-4}$) & 10 fb$^{-1}$ & 30 fb$^{-1}$ & 50 fb$^{-1}$ & 100 fb$^{-1}$ \\
\hline \hline
$f_{M0}/\Lambda^{4}$  & [-3.18;3.59] $\times 10^{2}$ & [-2.37;2.78] $\times 10^{2}$ & [-2.07;2.47] $\times 10^{2}$ & [-1.71;2.11] $\times 10^{2}$ \\
\hline
$f_{M1}/\Lambda^{4}$  & [-1.22;1.21] $\times 10^{3}$ & [-9.29;9.24]$ \times 10^{2}$  & [-8.18;8.13] $ \times 10^{2}$ & [-6.88;6.83] $ \times 10^{2}$   \\
\hline
$f_{M2}/\Lambda^{4}$  & [-0.56;0.47] $\times 10^{2}$ & [-0.44;0.35]$ \times 10^{2}$  & [-0.39;0.30] $ \times 10^{2}$ & [-0.34;0.25]$\times 10^{2}$   \\
\hline
$f_{M3}/\Lambda^{4}$  & [-1.85;1.86] $\times 10^{2}$ & [-1.40;1.42]$ \times 10^{2}$  & [-1.23;1.25] $ \times 10^{2}$ & [-1.04;1.05]$\times 10^{2}$   \\
\hline
$f_{M4}/\Lambda^{4}$  & [-1.77;1.97] $\times 10^{2}$ & [-1.33;1.52]$ \times 10^{2}$  & [-1.16;1.35] $ \times 10^{2}$ & [-0.96;1.15]$\times 10^{2}$   \\
\hline
$f_{M5}/\Lambda^{4}$  & [-6.84;6.65] $\times 10^{2}$ & [-5.22;5.03]$ \times 10^{2}$  & [-4.61;4.41] $ \times 10^{2}$ & [-3.89;3.70]$\times 10^{2}$   \\
\hline
$f_{M7}/\Lambda^{4}$  & [-2.39;2.49]$\times 10^{3}$    & [-1.81;1.90]$\times 10^{3}$   & [-1.59;1.68]$\times 10^{3}$   & [-1.33;1.42]$\times 10^{3}$    \\
\hline
$f_{T0}/\Lambda^{4}$  &  [-1.27;1.46] $\times 10^{1}$ & [-0.94;1.13] $\times 10^{1}$ & [-0.82;1.01] $\times 10^{1}$ & [-0.67;0.87] $\times 10^{1}$ \\
\hline
$f_{T1}/\Lambda^{4}$  & [-5.90;5.59] $\times 10^{1}$  & [-4.52;4.21] $\times 10^{1}$ & [-4.00;3.69] $\times 10^{1}$ & [-3.39;3.07] $\times 10^{1}$  \\
\hline
$f_{T2}/\Lambda^{4}$  &  [-5.09;5.56] $\times 10^{1}$ & [-3.81;4.28] $\times 10^{1}$ & [-3.33;3.80] $\times 10^{1}$ & [-2.77;3.24] $\times 10^{1}$ \\
\hline
$f_{T5}/\Lambda^{4}$  & [-4.39;3.91]  & [-3.40;2.91] & [-3.03;2.54] & [-2.59;2.10]  \\
\hline
$f_{T6}/\Lambda^{4}$  & [-1.84;1.66] $\times 10^{1}$  & [-1.42;1.24] $\times 10^{1}$ & [-1.26;1.08] $\times 10^{1}$ & [-1.08;0.90] $\times 10^{1}$ \\
\hline
$f_{T7}/\Lambda^{4}$  & [-1.61;1.64] $\times 10^{1}$  & [-1.22;1.25] $\times 10^{1}$ & [-1.08;1.10] $\times 10^{1}$ & [-0.90;0.93] $\times 10^{1}$ \\
\hline
\end{tabular}
\end{table}

\begin{table}
\caption{The limits obtained on the anomalous quartic $WW\gamma\gamma$ couplings at $95\%$ Confidence Level through the process $ep \rightarrow \nu_{e}\gamma \gamma j$ at 7.07 TeV FCC-he with integrated luminosities of 100, 300, 500, 1000 fb$^{-1}$.}

\begin{tabular}{|c|c|c|c|c|c|}\hline
Couplings (TeV$^{-4}$) & 100 fb$^{-1}$ & 300 fb$^{-1}$ & 500 fb$^{-1}$ & 1000 fb$^{-1}$ \\
\hline \hline
$f_{M0}/\Lambda^{4}$  & [-1.75;1.79] $\times 10^{1}$ & [-1.33;1.36] $\times 10^{1}$ & [-1.17;1.20] $\times 10^{1}$ & [-0.98;1.01] $\times 10^{1}$ \\
\hline
$f_{M1}/\Lambda^{4}$  & [-6.69;6.11] $\times 10^{1}$ & [-5.16;4.58]$ \times 10^{1}$  & [-4.57;4.00] $ \times 10^{1}$ & [-3.90;3.32] $ \times 10^{1}$   \\
\hline
$f_{M2}/\Lambda^{4}$  & [-2.80;2.60] & [-2.15;1.95]  & [-1.91;1.70] & [-1.62;1.42]  \\
\hline
$f_{M3}/\Lambda^{4}$  & [-9.50;10.01] & [-7.15;7.72]  & [-6.26;6.83] & [-5.22;5.79]   \\
\hline
$f_{M4}/\Lambda^{4}$  & [-9.61;9.91] & [-7.27;7.57]  & [-6.38;6.68] & [-5.34;5.64]   \\
\hline
$f_{M5}/\Lambda^{4}$  & [-3.58;3.50] $\times 10^{1}$ & [-2.74;2.65]$ \times 10^{1}$  & [-2.42;2.33] $ \times 10^{1}$ & [-2.04;1.95]$\times 10^{1}$   \\
\hline
$f_{M7}/\Lambda^{4}$  & [-1.27;1.30]$\times 10^{2}$    & [-9.58;9.94]$\times 10^{1}$   & [-8.41;8.77]$\times 10^{1}$   & [-7.04;7.40]$\times 10^{1}$    \\
\hline
$f_{T0}/\Lambda^{4}$  &  [-2.62;2.95] $\times 10^{-1}$ & [-1.96;2.29] $\times 10^{-1}$ & [-1.70;2.03] $\times 10^{-1}$ & [-1.41;1.74] $\times 10^{-1}$ \\
\hline
$f_{T1}/\Lambda^{4}$  & [-1.28;1.23]  & [-9.76;9.29] $\times 10^{-1}$ & [-8.62;8.15] $\times 10^{-1}$ & [-7.29;6.82] $\times 10^{-1}$  \\
\hline
$f_{T2}/\Lambda^{4}$  &  [-1.05;1.12] & [-0.79;0.86] & [-0.69;0.76] & [-0.58;0.65] \\
\hline
$f_{T5}/\Lambda^{4}$  & [-1.60;1.41] $\times 10^{-1}$  & [-1.24;1.05] $\times 10^{-1}$ & [-1.10;0.91]  $\times 10^{-1}$ & [-0.94;0.76]  $\times 10^{-1}$  \\
\hline
$f_{T6}/\Lambda^{4}$  & [-0.40;0.36]  & [-0.31;0.27] & [-0.28;0.23] & [-0.24;0.19] \\
\hline
$f_{T7}/\Lambda^{4}$  & [-3.29;3.30] $\times 10^{-1}$  & [-2.50;2.51] $\times 10^{-1}$ & [-2.20;2.21] $\times 10^{-1}$ & [-1.85;1.86] $\times 10^{-1}$ \\
\hline
\end{tabular}
\end{table}

\begin{table}
\caption{The same as Table VI but for 10 TeV FCC-he.}
\begin{tabular}{|c|c|c|c|c|c|}\hline
Couplings (TeV$^{-4}$) & 100 fb$^{-1}$ & 300 fb$^{-1}$ & 500 fb$^{-1}$ & 1000 fb$^{-1}$ \\
\hline \hline
$f_{M0}/\Lambda^{4}$  & [-5.76;6.13] & [-4.33;4.70] & [-3.79;4.16] & [-3.16;3.53] \\
\hline
$f_{M1}/\Lambda^{4}$  & [-2.24;2.12] $\times 10^{1}$ & [-1.72;1.60]$ \times 10^{1}$  & [-1.52;1.40] $ \times 10^{1}$ & [-1.29;1.17] $ \times 10^{1}$   \\
\hline
$f_{M2}/\Lambda^{4}$  & [-0.94;0.87] & [-0.72;0.65]  & [-0.64;0.57] & [-0.54;0.47]   \\
\hline
$f_{M3}/\Lambda^{4}$  & [-3.46;3.19] & [-2.66;2.39]  & [-2.36;2.09] & [-2.01;1.74]   \\
\hline
$f_{M4}/\Lambda^{4}$  & [-3.19;3.39] & [-2.41;2.60]  & [-2.11;2.30] & [-1.76;1.95]   \\
\hline
$f_{M5}/\Lambda^{4}$  & [-1.26;1.16] $\times 10^{1}$ & [-0.97;0.87]$ \times 10^{1}$  & [-0.86;0.76] $ \times 10^{1}$ & [-0.73;0.63]$\times 10^{1}$   \\
\hline
$f_{M7}/\Lambda^{4}$  & [-4.18;4.54]$\times 10^{1}$    & [-3.13;3.49]$\times 10^{1}$   & [-2.74;3.10]$\times 10^{1}$   & [-2.28;2.64]$\times 10^{1}$    \\
\hline
$f_{T0}/\Lambda^{4}$  &  [-7.62;7.65] $\times 10^{-2}$ & [-5.79;5.81] $\times 10^{-2}$ & [-5.09;5.12] $\times 10^{-2}$ & [-4.28;4.31] $\times 10^{-2}$ \\
\hline
$f_{T1}/\Lambda^{4}$  & [-3.34;3.30] $\times 10^{-1}$ & [-2.55;2.50] $\times 10^{-1}$ & [-2.24;2.20] $\times 10^{-1}$ & [-1.89;1.85] $\times 10^{-1}$ \\
\hline
$f_{T2}/\Lambda^{4}$  &  [-2.95;3.09] $\times 10^{-1}$ & [-2.23;2.36] $\times 10^{-1}$ & [-1.95;2.09] $\times 10^{-1}$ & [-1.63;1.77] $\times 10^{-1}$ \\
\hline
$f_{T5}/\Lambda^{4}$  & [-4.24;4.08] $\times 10^{-2}$  & [-3.24;3.08] $\times 10^{-2}$ & [-2.87;2.70] $\times 10^{-2}$ & [-2.42;2.26] $\times 10^{-2}$  \\
\hline
$f_{T6}/\Lambda^{4}$  & [-1.04;1.00] $\times 10^{-1}$  & [-7.91;7.54] $\times 10^{-2}$ & [-6.99;6.62] $\times 10^{-2}$ & [-5.91;5.54] $\times 10^{-2}$ \\
\hline
$f_{T7}/\Lambda^{4}$  & [-0.86;0.98] $\times 10^{-1}$ & [-0.64;0.76] $\times 10^{-1}$ & [-0.55;0.68] $\times 10^{-1}$ & [-0.46;0.58] $\times 10^{-1}$ \\
\hline
\end{tabular}
\end{table}

\end{document}